\documentclass[fleqn,usenatbib]{mnras}

\usepackage{newtxtext,newtxmath}

\usepackage[T1]{fontenc}

\DeclareRobustCommand{\VAN}[3]{#2}
\let\VANthebibliography\thebibliography
\def\thebibliography{\DeclareRobustCommand{\VAN}[3]{##3}\VANthebibliography}

\usepackage{graphicx}	        
\usepackage{amsmath}	        
\usepackage{multirow}           
\usepackage{ulem}               
\usepackage[dvipsnames]{xcolor} 
\usepackage{soul}

\mathchardef\shorthyphen="2D                                            
\newcommand{\mSun}{\; {\rm M_\odot}}                            
\newcommand{\mJup}{\; {\rm M_{Jup}}}                            
\newcommand{\mEarth}{\; {\rm M_\oplus}}                         
\newcommand{\au}{\; {\rm au}}                                           
\newcommand{\m}{\; {\rm m}}                                                     
\newcommand{\cm}{\; {\rm cm}}                                                     
\newcommand{\mm}{\; {\rm mm}}                                                     
\newcommand{\km}{\; {\rm km}}                                           
\newcommand{\um}{\; {\rm \mu m}}                                        
\newcommand{\yr}{\; {\rm yr}}                                         
\newcommand{\myr}{\; {\rm Myr}}                                         
\newcommand{\gyr}{\; {\rm Gyr}}                                         
\newcommand{\kmPerS}{\; {\rm km \; s^{-1}}}                     
\newcommand{\gPerCmCubed}{\; {\rm g \; cm^{-3}}}        
\newcommand{\percent}{\; {\rm per \; cent}}                     
\newcommand{\rad}{\; {\rm radians} \;}                                         
\newcommand{\HD}{{\rm HD} \;}                                           
\newcommand{\HR}{{\rm HR} \;}                                           

\newcommand{\rHill}{r_{\rm H}}
\newcommand{\SigmaR}{\Sigma(r)}
\newcommand{\rI}{r_{\rm i}}

\newcommand{\rO}{r_{\rm o}}
\newcommand{\sigmaRI}{\sigma_{{\rm i}}}
\newcommand{\sigmaRO}{\sigma_{{\rm o}}}
\newcommand{\alphaR}{\alpha}

\newcommand{\rIZero}{r_{\rm i, 0}}
\newcommand{\aP}{a_{\rm p}}
\newcommand{\eP}{e_{\rm p}}
\newcommand{\mP}{m_{\rm p}}
\newcommand{\eMaxZero}{e_{\rm max, 0}}
\newcommand{\eRMSI}{e_{\rm i, rms}}
\newcommand{\eRMSIFromP}{e_{\rm i, rms, p}}
\newcommand{\eRMSIZero}{e_{\rm i, rms, 0}}
\newcommand{\deltaA}{\delta a}
\newcommand{\mPOverMStar}{\mP/m_*}
\newcommand{\tauMax}{\tau_{\rm max}}
\newcommand{\tauB}{\tau_{\rm b}}
\newcommand{\sMax}{s_{\rm max}}
\newcommand{\sMin}{s_{\rm min}}
\newcommand{\sB}{{s_{\rm b}}}
\newcommand{\sD}{{s_{\rm dust}}}
\newcommand{\xM}{{x_{\rm m}}}
\newcommand{\qP}{{q_{\rm p}}}
\newcommand{\qG}{{q_{\rm g}}}
\newcommand{\qS}{{q_{\rm s}}}
\newcommand{\bS}{{b_{\rm s}}}
\newcommand{\bG}{{b_{\rm g}}}
\newcommand{\xC}{X_{\rm c}}
\newcommand{\qDStar}{Q_{\rm D}^*}
\newcommand{\rT}{r_{\rm t}}
\newcommand{\pI}{{p_{\rm i}}}
\newcommand{\pO}{{p_{\rm o}}}
\newcommand{\eMaxSec}{{e_{\rm max, sec}}}
\newcommand{\tDiff}{{t_{\rm diff}}}
\newcommand{\aSec}{{a_{\rm res}}}

\long\def\symbolfootnote[#1]#2{\begingroup%
\def\thefootnote{\fnsymbol{footnote}}\footnote[#1]{#2}\endgroup} 
\def\aj{AJ}
\def\araa{ARA\&A}
\def\apj{ApJ}
\def\apjl{ApJ}
\def\apss{Ap\&SS}
\def\aap{A\&A}
\def\icarus{Icarus}
\def\mnras{MNRAS}
\def\nat{Nature}

\title[The effect of planets on debris-disc edges]{The effect of sculpting planets on the steepness of debris-disc inner edges}

\author[T. D. Pearce et al.]{Tim D. Pearce$^{1,2}$\thanks{E-mail: tim.pearce@warwick.ac.uk},
Alexander V. Krivov$^{2}$,
Antranik A. Sefilian$^{2, 3}$,
Marija R. Jankovic$^4$,
Torsten L\"{o}hne$^2$,\newauthor
Tobias Morgner$^2$,
Mark C. Wyatt$^5$,
Mark Booth$^{6,2}$,
Sebastian Marino$^7$\\
$^1$Department of Physics, University of Warwick, Gibbet Hill Road, Coventry CV4 7AL, UK\\
$^2$Astrophysikalisches Institut und Universit\"{a}tssternwarte, Friedrich-Schiller-Universit\"{a}t Jena, Schillerg\"{a}{\ss}chen 2-3, D-07745 Jena, Germany\\
$^3$Center for Advanced Mathematical Sciences, American University of Beirut, PO Box 11-0236, Riad El-Solh, Beirut 11097 2020, Lebanon\\
$^4$Institute of Physics Belgrade, University of Belgrade, Pregrevica 118, 11080 Belgrade, Serbia\\
$^5$Institute of Astronomy, University of Cambridge, Madingley Road, Cambridge CB3 0HA, UK\\ 
$^6$UK Astronomy Technology Centre, Royal Observatory Edinburgh, Blackford Hill, Edinburgh EH9 3HJ, UK\\
$^7$Department of Physics and Astronomy, University of Exeter, Stocker Road, Exeter, EX4 4QL, UK\\
}

\date{Accepted XXX. Received YYY; in original form ZZZ}

\pubyear{2023}

\begin{document}
\label{firstpage}
\pagerange{\pageref{firstpage}--\pageref{lastpage}}
\maketitle

\begin{abstract}
Debris discs are our best means to probe the outer regions of planetary systems. Many studies assume that planets lie at the inner edges of debris discs, akin to Neptune and the Kuiper Belt, and use the disc morphologies to constrain those otherwise-undetectable planets. However, this produces a degeneracy in planet mass and semimajor axis. We investigate the effect of a sculpting planet on the radial surface-density profile at the disc inner edge, and show that this degeneracy can be broken by considering the steepness of the edge profile. Like previous studies, we show that a planet on a circular orbit ejects unstable debris and excites surviving material through mean-motion resonances. For a non-migrating, circular-orbit planet, in the case where collisions are negligible, the steepness of the disc inner edge depends on the planet-to-star mass ratio and the initial-disc excitation level.
We provide a simple analytic model to infer planet properties from the steepness of ALMA-resolved disc edges. We also perform a collisional analysis, showing that a purely planet-sculpted disc would be distinguishable from a purely collisional disc and that, whilst collisions flatten planet-sculpted edges, they are unlikely to fully erase a planet's signature. Finally, we apply our results to ALMA-resolved debris discs and show that, whilst many inner edges are \textit{too steep} to be explained by collisions alone, they are \textit{too flat} to arise through completed sculpting by non-migrating, circular-orbit planets. We discuss implications of this for the architectures, histories and dynamics in the outer regions of planetary systems. 
\end{abstract}

\begin{keywords}
planet-disc interactions -- planets and satellites: dynamical evolution and stability -- circumstellar matter
\end{keywords}

\section{Introduction}
\label{sec: intro}

Debris discs, like the Solar System's Asteroid and Kuiper Belts, are circumstellar populations of sub-planet-mass objects \citep{Wyatt2008}. These objects undergo destructive collisions that release observable dust, and such dust is detected around \mbox{20\%} of main-sequence stars \citep{Hughes2018, Wyatt2020}. Extrasolar debris discs can be resolved in scattered light and thermal emission (e.g. \citealt{Esposito2020, Lovell2021, Booth2023}), and are invaluable probes of processes in the outer regions of planetary systems; modern planet-detection techniques are insensitive to mid-sized planets orbiting at 10s or 100s of au from stars, but such planets can be inferred from their influence on observed debris (e.g. \citealt{Mouillet1997, Wyatt1999, Quillen2006Fom, Pearce2014, Sefilian2021, Stuber2023}). The launch of the \textit{James Webb Space Telescope} (\textit{JWST}) has prompted renewed interest in such hypothetical planets, because if they exist, then many should be detectable by this facility for the first time \citep{Pearce2022ISPY}.

Many theoretical studies have used the shapes and locations of debris discs to infer the properties of unseen planets. If such planets orbit just interior to debris discs, and are responsible for sculpting debris-disc inner edges, then the shapes and locations of these edges can be used to infer the planets' minimum-possible mass, maximum semimajor axis and minimum eccentricity (e.g. \citealt{Pearce2014, Faramaz2019, Pearce2022ISPY}). However, this approach suffers from several key unknowns. First, whilst Neptune dominates the Kuiper Belt's inner edge \citep{Malhotra1993}, it is unclear whether planets actually \textit{do} reside just interior to extrasolar debris discs. It could be that the disc shapes and locations are instead set by other, non-planetary processes, such as debris-debris collisions or system formation. Second, even if exoplanets do sculpt debris-disc edges, it is unclear whether they do so on fixed orbits (as often assumed in planet-disc-interaction studies), or whether they migrate over time. Third, even if planetary sculpting occurs, it is unclear whether this process has finished in observed discs (as often assumed), or whether disc edges are still evolving under planet-debris interactions. These distinctions have significant implications for the morphology of debris discs, and the masses and locations of any responsible planets today. Such questions are difficult to answer using only debris-disc shapes and inner-edge locations.

However, we can gain more insight by also considering the \textit{radial surface-density profile} of debris-disc edges. If an edge were sculpted by a non-migrating planet, then this profile would have a characteristic steepness that depends on planet mass \citep{Mustill2012, Rodigas2014, Quillen2006Fom, Chiang2009}. Conversely, if the edge were set by non-planetary processes like collisions, then its steepness would be different \citep{Wyatt2007, Lohne2008, Geiler2017, ImazBlanco2023}. Similarly, the profiles of edges sculpted by migrating planets would be distinct from those due to non-migrating planets \citep{Friebe2022}, and edges set by combinations of both planetary and non-planetary effects would also have different profiles \citep{Thebault2012, Nesvold2015}. The profiles of debris-disc edges are therefore powerful tools for probing the architectures, histories and processes in the outer regions of planetary systems.

Recently, several studies used ALMA to resolve the radial profiles of debris-disc edges (e.g. \citealt{Lovell2021, Faramaz2021, Marino2021}). These observations trace millimetre-sized grains that are unaffected by radiation forces \citep{Hughes2018}, so offer our best means to probe the distribution of larger, unseen planetesimals in the discs. Such studies show that some debris discs have flatter inner edges that are consistent with pure collisional evolution, whilst others have steeper profiles potentially indicative of planetary sculpting \citep{ImazBlanco2023}. Our aim in this paper is to thoroughly investigate the dynamical effects of sculpting planets on the steepness of debris-disc inner edges, and establish whether any of the observed ALMA steepnesses are consistent with planetary sculpting. We will show that, whilst many inner edges are \textit{too steep} to be explained through collisions alone, they are \textit{too flat} to be purely the result of completed sculpting by non-migrating, circular-orbit planets. This potentially implies that different processes, or combinations of processes, are actually operating in these discs.

Several previous studies also explored the effect of sculpting planets on the steepness of debris-disc inner edges (e.g. \citealt{Quillen2006Fom, QuillenFaber2006, Chiang2009, Mustill2012, Rodigas2014, Nesvold2015, Regaly2018}). Our study differs from these in several ways. First, since we do not know debris-disc masses or the sizes of the largest planetesimals \citep{KrivovWyatt2021}, both of which are critical for quantifying collisional evolution, we choose to decouple the dynamical and collisional modelling in this paper. Instead, we initially model the interaction between a planet and a collisionless disc, to isolate the effect of a planet on the edge steepness. We then apply a collisional model, to demonstrate how collisions would change the steepness from the planet-only regime. Second, we explore a very broad parameter space in both the dynamical and collisional simulations, to provide general predictive models that are valid across a wide range of scenarios. Third, we parameterise our results in a form that directly relates to those fitted to ALMA data, to facilitate the application of our results to upcoming observations. Finally, we apply our predictive model to recent ALMA observations of debris discs.

The layout of this paper is as follows. Section \ref{sec: simulations} describes our $n$-body simulations of a circular-orbit planet and a collisionless debris disc. Section \ref{sec: simpleAnalyticModel} provides a predictive model relating the steepness of a disc's edge to the parameters of a circular-orbit planet, and specifically Section \ref{subsec: modelExampleFitStepByStepMethod} gives a step-by-step method demonstrating how to infer an unseen sculpting planet from a debris disc's inner edge. Section \ref{sec: collisions} addresses how collisions would affect planet-sculpted edges. In Section \ref{sec: applicationToObservedDiscs} we apply our results to observed debris discs, and in Section \ref{sec: discussion} we discuss our results (including what the observed-disc profiles may be telling us about planetary systems). We conclude in Section \ref{sec: conclusions}.

\section{N-body simulations}
\label{sec: simulations}

We use a large suite of $n$-body simulations to explore the dynamical interaction between a circular-orbit planet and a collisionless debris disc. The effects of debris-debris collisions will be assessed later (Section \ref{sec: collisions}). We describe the circular-planet simulation setup in Section \ref{subsec: simulationsSetup}, discuss two example simulations in Section \ref{subsec: simulationsExamples}, and present the surface-density profiles of all simulated discs in Section \ref{subsec: simulationsSDFit}. We measure the debris-excitation level at the simulated inner edges in Section \ref{subsec: simulationsSDEccentricityOfInnerEdge}, and show this to arise through MMR interactions with the circular-orbit planet. In addition, we also extend the $n$-body analyses to eccentric planets in \mbox{Appendix \ref{app: eccentricPlanets}}.

\subsection{Setup of $n$-body simulations}
\label{subsec: simulationsSetup}

\begin{figure}
	\includegraphics[width=8cm]{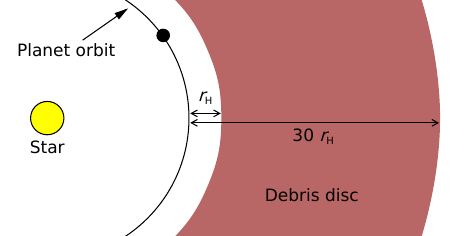}
    \caption{Initial setup of our $n$-body simulations with a circular-orbit planet. A single planet (black circle) is initialised just interior to a massless debris disc, which spans 1 to 30 Hill radii exterior to the planet orbit.}
    \label{fig: setupDiagram}
\end{figure}

The initial setup of our $n$-body simulations with circular-orbit planets is shown on Figure \ref{fig: setupDiagram}. Each simulation comprises one star, one planet, and a disc of 20,000 massless debris particles. We run over 300 simulations, each with a different combination of star mass, planet mass, planet semimajor axis and initial-disc excitation level.

Each debris particle is initialised with a semimajor axis between 1 and 30 Hill radii ($\rHill$) exterior to the planet;

\begin{equation}
    \rHill \equiv \aP \left(\frac{m_{\rm p}} {3 m_* }\right)^{1/3},
    \label{eq: rHillCircular}
\end{equation}

\noindent where $\aP$ is the planet semimajor axis and $m_{\rm p}$ and $m_*$ are the planet and star masses respectively. The inner edge of 1 Hill radius ensures that Trojans are omitted, which would otherwise bias the axisymmetric surface-density profiles that we fit later (we discuss Trojans in Section \ref{subsec: discussionTrojans}). The outer edge of 30 Hill radii ensures that the outer edge does not affect the inner-edge profile; a circular-orbit planet is expected to scatter all non-resonant material originating within approximately 3 Hill radii of its orbit (\citealt{Gladman1993, Ida2000, Kirsh2009, Malhotra2021, Friebe2022}), so setting the outer edge at $30\rHill$ ensures that this is well beyond the inner region. Debris semimajor axes are drawn such that the initial surface-density distribution goes as approximately ${r^{-1.5}}$ (where $r$ is stellocentric distance), akin to the Minimum-Mass Solar Nebula (MMSN; \citealt{Weidenschilling1977, Hayashi1981}). Each debris particle has an initial eccentricity uniformly drawn between 0 and a maximum value ${\eMaxZero}$, and an initial inclination (relative to the planet's orbital plane) uniformly drawn between 0 and ${\eMaxZero/2 \rad}$. Each particle's initial longitude of ascending node, argument of pericentre and mean anomaly are uniformly drawn between 0 and ${360^\circ}$.

To ensure that scattering is essentially complete by the end of the simulations, we set each simulation end time depending on the diffusion timescale $\tDiff$. The value $\tDiff$ characterises the scattering timescale, and is roughly the time taken for a planet to significantly scatter or eject ${70\percent}$ of material originating within 3 Hill radii of its orbit (Costa, Pearce \& Krivov, submitted). The diffusion time for a planet acting on a body with semimajor axis $a$ is 

\begin{equation}
    \tDiff \approx 0.01 T_{\rm p} \left(\frac{\aP}{a}\right)^{1/2} \left(\frac{m_{\rm p}}{m_*}\right)^{-2},
    \label{eq: diffusionTime}
\end{equation}

\noindent where $T_{\rm p}$ is the planet's orbital period \citep{tremaine1993}. Equivalently,

\begin{equation}
    \tDiff \approx 1.1 \times 10^4\yr \;
    \left(\frac{\aP}{\au}\right)^2
    \left(\frac{a}{\au}\right)^{-1/2}
    \left(\frac{m_*}{\mSun}\right)^{3/2}
    \left(\frac{\mP}{\mJup}\right)^{-2}.
    \label{eq: diffusionTimeUnits}
\end{equation}

\noindent We run each simulation for at least ${10 \tDiff}$ calculated at ${a=\aP}$, to ensure that scattering is essentially complete by the end. Simulations are run in {\sc rebound} \citep{Rein2012Rebound} using {\sc whfast} (a symplectic Wisdom-Holman integrator; \citealt{Rein2015,Wisdom1991}), with a timestep of ${1\percent}$ of the planet's orbital period\footnote{{\sc rebound} does not define several orbital parameters in the case of a perfectly circular orbit. To ensure correct behaviour, for our `circular-planet' simulations we actually implement a planet eccentricity of ${10^{-4}}$.}. Simulations are conducted and analysed in the centre-of-mass frame.

We run simulations with star masses of \mbox{1 to ${2 \mSun}$}, planet-to-star mass ratios of ${3 \times 10^{-5}}$ to $10^{-1}$, planet semimajor axes of 1 to ${100\au}$, and maximum initial debris eccentricities of ${\eMaxZero=0.001}$ to 0.3. In addition, we also re-weight debris particles in post-simulation analyses, so we can test different initial surface-density profiles; we consider initial surface-density profiles between ${r^0}$ and ${r^{-1.5}}$. We assume the planet and debris to be point-like particles; this lets us apply simple scaling laws to our results, but also means that the potential removal of debris via accretion rather than ejection is ignored in our simulations (see \citealt{Morrison2015}).

\subsection{Example simulation results}
\label{subsec: simulationsExamples}

\begin{figure*}
    \includegraphics[width=17cm]{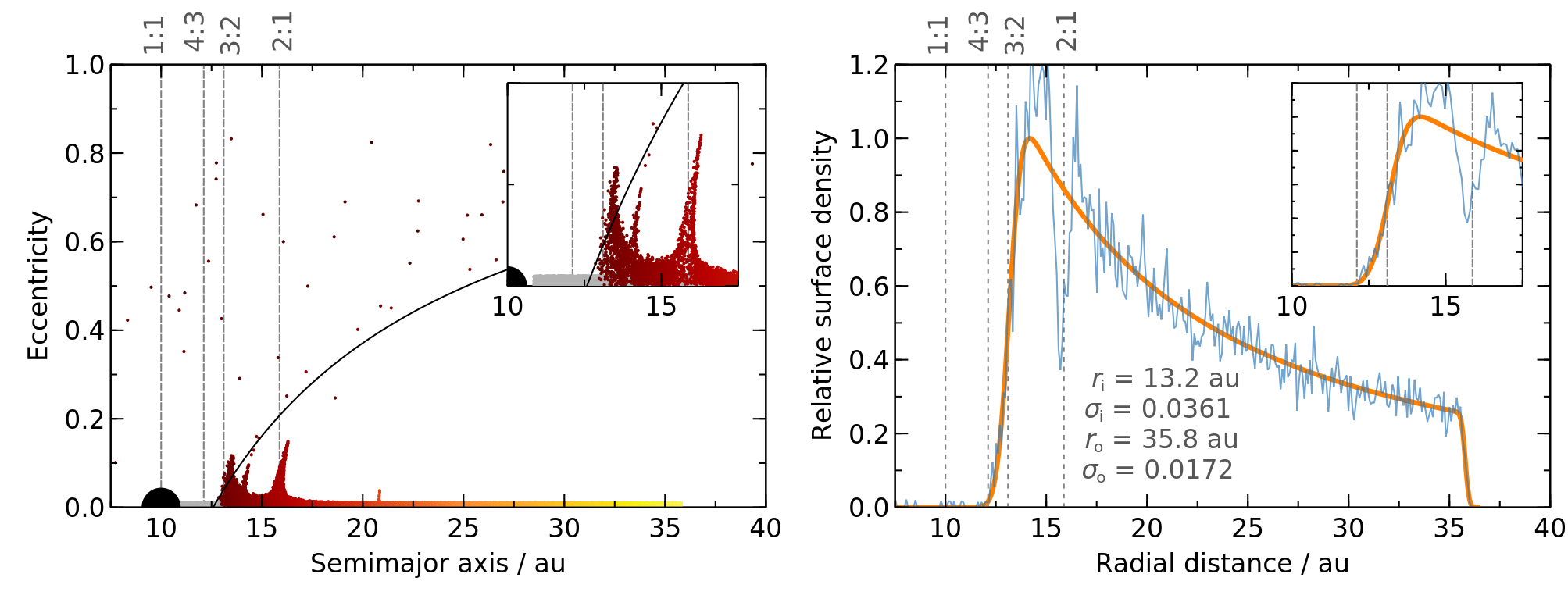}
    \caption{Simulation of a circular-orbit planet interacting with an exterior debris disc, for an initially weakly excited disc. The simulation is of a solar-mass star, a ${2 \mJup}$ planet on a circular orbit at ${10\au}$ (${m_{\rm p}/m_* = 2\times10^{-3}}$), and 20,000 massless debris particles initially spanning 1 to 30 Hill radii exterior to the planet, with an initial surface-density profile of approximately ${r^{-1.5}}$. Initial debris eccentricities were each uniformly drawn between 0 and 0.01, and the plot shows the final state after ${8.7 \times 10^5 \yr}$ (10 diffusion timescales). The planet clears nearby debris through scattering, and excites the eccentricities of surviving material through MMRs. Left panel: semimajor axes and eccentricities of the planet (large circle) and debris (small points, coloured by initial semimajor axis). Grey points show debris at the start of the simulation. Solid black lines are the minimum eccentricities required for debris to come within 3 Hill radii of the planet, and dashed grey lines are nominal MMR locations. Right panel: radial surface-density profile of the disc, $\SigmaR$. The thin blue line shows the simulation data, and the thick orange line is the fit (Equation \ref{eq: fittedProfile}). Insets show the inner-edge region in more detail. The fitted model parameters are shown on the right panel; $\alphaR$ was fixed to 1.5 in the fit.}
    \label{fig: lowESurfaceDensities}
\end{figure*}

\begin{figure*}
    \includegraphics[width=17cm]{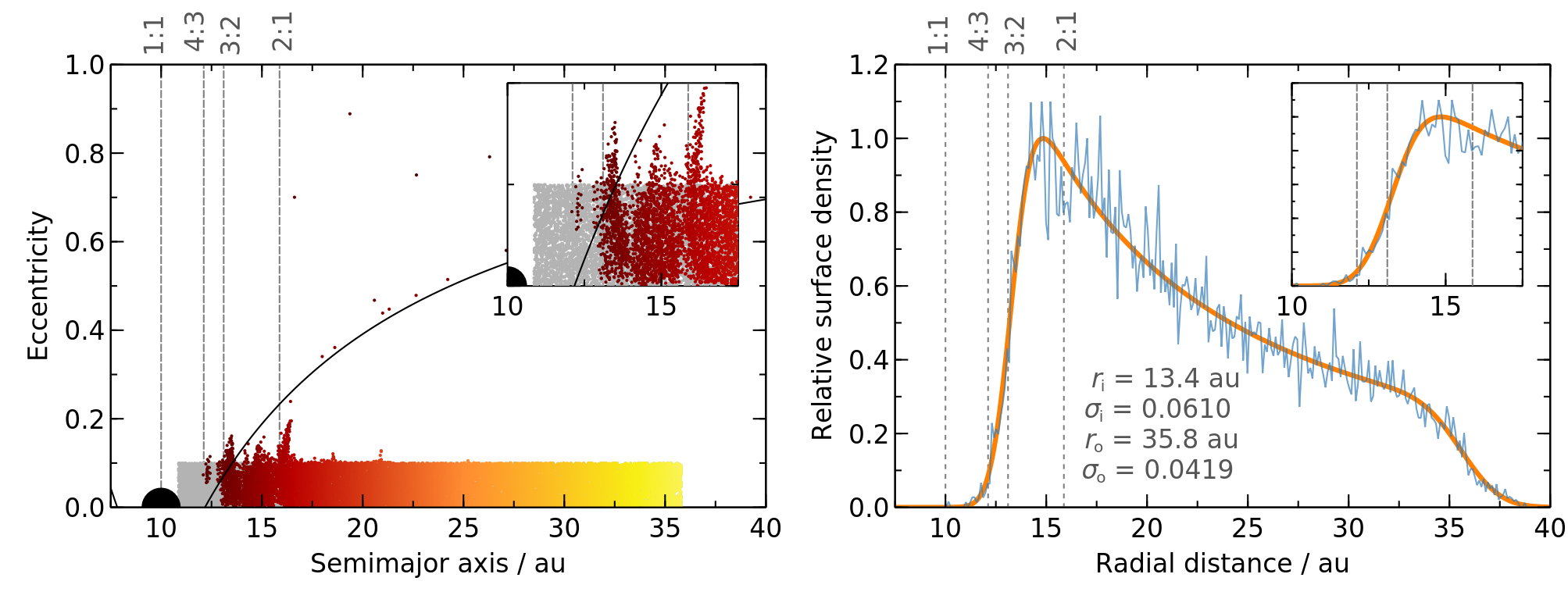}
    \caption{Simulation with the same setup as that on Figure \ref{fig: lowESurfaceDensities}, except that the debris disc is initially more excited (maximum initial eccentricities of 0.1, rather than the value of 0.01 on Figure \ref{fig: lowESurfaceDensities}). The planet still excites debris through MMRs (left panel), but these populations are less significant against the background excitation level of the disc. The resulting inner edge is flatter than in the low-initial-excitation case.}
    \label{fig: highESurfaceDensities}
\end{figure*}

Figures \ref{fig: lowESurfaceDensities} and \ref{fig: highESurfaceDensities} show two example simulations after 10 diffusion timescales. Both have a solar-mass star and a ${2 \mJup}$ planet (${m_{\rm p}/m_* = 2\times10^{-3}}$), which is on a circular orbit at ${10\au}$. The simulation on Figure \ref{fig: lowESurfaceDensities} starts with an initially unexcited debris disc (${\eMaxZero=0.01}$), and that on Figure \ref{fig: highESurfaceDensities} has the same setup but an initially excited disc (${\eMaxZero=0.1}$). The main qualitative features of these simulations are typical of all our runs.

The left panels of Figures \ref{fig: lowESurfaceDensities} and \ref{fig: highESurfaceDensities} show the semimajor axes and eccentricities of the simulated bodies, which demonstrate the two main effects of a circular-orbit planet on debris. The planet: 

\begin{enumerate}
    \item scatters and ejects most debris coming within \mbox{${\sim 3}$ Hill} radii of its orbit; this clears the region above the solid black lines on the left panels.

    \item excites the eccentricities of surviving debris via mean-motion resonances (MMRs), particularly near the inner edge of the sculpted disc.
\end{enumerate}

The role of MMRs in exciting debris eccentricity is clearest in the initially unexcited-disc simulation (Figure \ref{fig: lowESurfaceDensities}, left panel). Here, populations of debris in the 2:1 and 3:2 MMRs are clearly visible. For the initially excited disc (Figure \ref{fig: highESurfaceDensities}), these MMR populations are similar to those in the initially unexcited simulation, only now they are less distinct due to the higher intrinsic eccentricity of the disc. The evolution of debris inclination is much less significant than eccentricity; there is some very small inclination excitation at specific resonances in these simulations, but the disc still remains thin across its entire width.

The right panels of Figures \ref{fig: lowESurfaceDensities} and \ref{fig: highESurfaceDensities} show the azimuthally averaged surface density profiles of the simulated debris discs (thin blue lines). The planet imposes a profile on the disc inner edge, whilst the central and outer regions retain their initial profiles. Individual, strong MMRs can also impose additional structure, such as the 2:1 MMR at ${15.9\au}$ on Figure \ref{fig: lowESurfaceDensities}; in this example that MMR does not affect the inner-edge fit, but it can do in other simulations. The thick orange lines are parametric fits to the surface-density profiles, as detailed in Section \ref{subsec: simulationsSDFit}.

\begin{figure}
    \includegraphics[width=7cm]{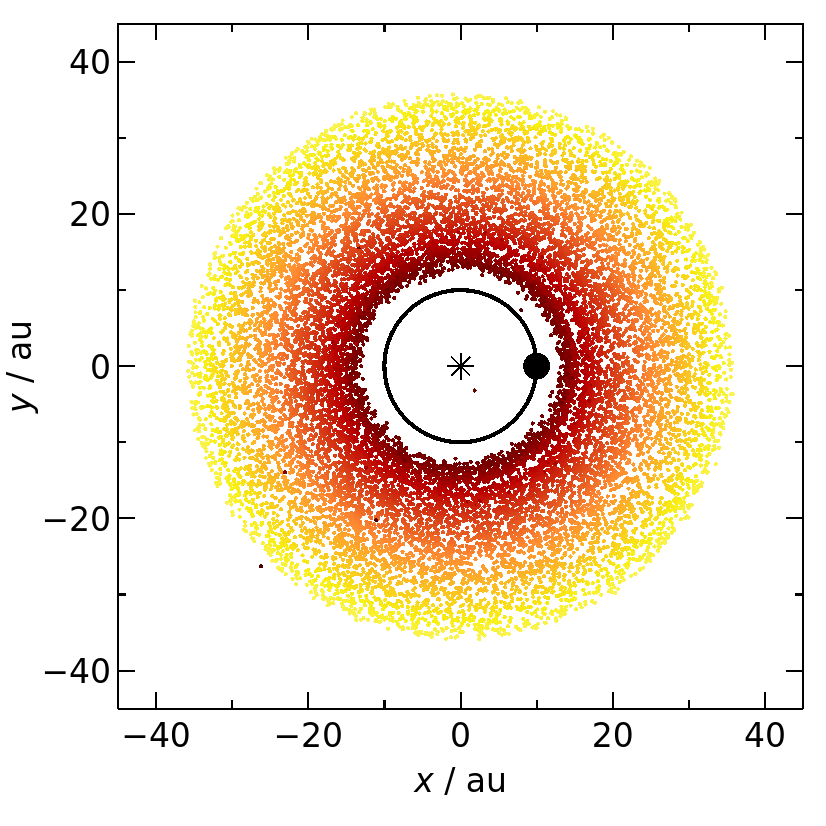}
    \caption{Positions of bodies from the simulation with a low-eccentricity disc (Figure \ref{fig: lowESurfaceDensities}). The asterisk, filled black circle and solid line denote the star, planet and planet orbit respectively, and small points are debris (coloured by initial semimajor axis). The disc is almost axisymmetric, with a slight asymmetry between the left and right sides due to MMRs. This causes a small difference in the inner-edge steepness between the two sides; the right side (aligned with the planet position) has ${\sigmaRI=0.0286}$, whilst the value for the opposite side is 0.0354.}
    \label{fig: lowESimPos}
\end{figure}

Figure \ref{fig: lowESimPos} shows the final positions of bodies in the initially unexcited-disc simulation (that on Figure \ref{fig: lowESurfaceDensities}). The disc is almost axisymmetric, with a slight asymmetry between the directions aligned and anti-aligned with the planet due to MMR structure (an effect noted by \citealt{Tabeshian2016, Tabeshian2017}). This leads to a small difference in the steepness of the inner-edge profile between the two sides of the disc, which we discuss further in Section \ref{subsec: discussionComparisonToPreviousStudiesGravity}. For the initially excited-disc simulation from Figure \ref{fig: highESurfaceDensities}, the asymmetry is less pronounced.

 
\subsection{Surface-density profile fitting}
\label{subsec: simulationsSDFit}

We aim to quantify how the properties of a sculpting planet affect the steepness and location of the debris disc's inner edge. To proceed, we use a parametric model to fit the radial surface-density profiles of debris in the $n$-body simulations, and examine how the model parameters change as functions of system properties. Section \ref{subsec: simulationsSDModel} describes the model, Section \ref{subsec: simulationsSDFitResultsAtLateTimes} the dependence of the fitted models on system parameters, and Section \ref{subsec: simulationsSDTimeEvolution} the timescale for a disc edge takes to reach its final state.

\subsubsection{Surface-density profile model}
\label{subsec: simulationsSDModel}

To quantify the disc surface-density profile, we follow the approach of \cite{Rafikov2023}. They show that, for a low-eccentricity disc with a sharp semimajor-axis cutoff at the outer edge, the radial surface-density profile around that edge can be characterised as\footnote{\cite{Rafikov2023} derived Equation \ref{eq: rafikov2023OuterEdgeProfile} analytically. \citet{Marino2021} fitted edges numerically using hyperbolic tangents instead, but the two profiles have similar forms.}

\begin{equation}
    \SigmaR \propto 1-{\rm erf}\left(\frac{r-\rO}{\sqrt{2} \sigmaRO \rO} \right).
    \label{eq: rafikov2023OuterEdgeProfile}
\end{equation}

\noindent Here ${\rO}$ characterises the radial location of the outer edge, $\sigmaRO$ is the `flatness' of the surface-density profile at that edge, and erf is the Gauss error function:

\begin{equation}
    {\rm erf}(z) \equiv \frac{2}{\sqrt{\pi}} \int^z_0 \exp({-t^2}) {\rm d}t.
    \label{eq: erfFunction}
\end{equation}

\noindent \mbox{Equation \ref{eq: rafikov2023OuterEdgeProfile}} is an `S-shape' profile centered on $\rO$, which is steeper for smaller $\sigmaRO$ and flatter for larger $\sigmaRO$. The value of $\sigmaRO$ is strongly linked to debris eccentricity; if debris has a sharp cutoff in semimajor axis, and its root-mean-square (rms) eccentricity is $e_{\rm rms}$, then ${\sigmaRO=e_{\rm rms}/\sqrt{2}}$ \citep{Rafikov2023}. This means that, for discs with sharp cutoffs in semimajor axis, lower rms eccentricities mean steeper edges (and smaller $\sigmaRO$). If the eccentricities are uniformly spread from 0 to $e_{\rm max}$, then ${e_{\rm rms} = e_{\rm max}/\sqrt{3}}$ and hence ${\sigmaRO = e_{\rm max}/\sqrt{6}}$. We will later show that a planet-sculpted disc does not have a sharp cutoff in semimajor axis at the inner edge, so the corresponding relationship between the edge profile and debris eccentricity is slightly modified, but the two remain strongly linked.

To parameterise the surface density across a whole disc, we use profiles similar to \mbox{Equation \ref{eq: rafikov2023OuterEdgeProfile}} at the two edges, combined with an $r^{-\alphaR}$ profile describing the surface density between the edges. By multiplying these three local profiles together, we arrive at the following model for the overall surface-density profile:

\begin{multline}
    \SigmaR = \frac{\Sigma_0}{2} \left[ 1-{\rm erf}\left(\frac{\rI - r}{\sqrt{2} \sigmaRI \rI} \right) \right] \left[ 1-{\rm erf}\left(\frac{r-\rO}{\sqrt{2} \sigmaRO \rO} \right) \right] \left(\frac{r}{\rI} \right)^{-\alphaR},
    \label{eq: fittedProfile}
\end{multline}

\noindent where subscripts ${\rm i}$ and ${\rm o}$ denote terms characterising the inner and outer edges respectively, and ${\Sigma_0}$ is the the surface density at ${r\approx\rI}$.

We numerically fit a radial profile of this form to each of our simulated discs, treating ${\Sigma_0}$, $\rI$, $\rO$, $\sigmaRI$, $\sigmaRO$ as free parameters. We typically fix $\alphaR$ to the initial surface-density index of the disc (i.e. ${\alphaR=1.5}$ for an initially MMSN profile). The fitting procedure is described in Appendix \ref{app: fittingProcedure}, and the orange lines on the right panels of Figures \ref{fig: lowESurfaceDensities} and \ref{fig: highESurfaceDensities} show the resulting fitted profiles for those simulations.

\subsubsection{Dependence of the final inner-edge profile on system parameters}
\label{subsec: simulationsSDFitResultsAtLateTimes}

Having fitted surface-density profiles to each of our $n$-body discs, we now assess how the profile of a planet-sculpted inner edge depends on system parameters. The inner-edge profiles are fully characterised by $\rI$ and $\sigmaRI$, and in this section we show how those fitted parameters vary with simulation setup.

\begin{figure}
	\includegraphics[width=8cm]{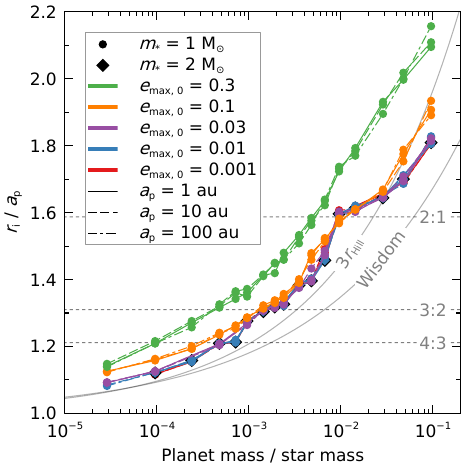}
    \caption{Location of the sculpted disc's inner edge $\rI$ (in terms of planet semimajor axis $\aP$), from fits to surface-density profiles at the end of our $n$-body simulations. Coloured points and lines show simulations with different initial setups, as denoted in the key. Grey solid lines denote different definitions of the outer edge of the chaotic zone; after the planet has ejected unstable debris, the disc's inner edge is typically just outside this chaotic zone. Dashed horizontal lines denote where $\rI$ coincides with the nominal location of a strong MMR; these resonances can significantly affect the edge location.}
    \label{fig: rIn_ap}
\end{figure}

Figure \ref{fig: rIn_ap} shows how the location of the inner edge of a planet-sculpted disc, $\rI$, scales with the planet-to-star mass ratio, planet semimajor axis, and the initial disc-excitation level. The location $\rI$ is just exterior to the `chaotic zone' around the planet's circular orbit, which can be defined as either \mbox{${\sim 3}$ Hill} radii or via the Wisdom overlap criterion \citep{Wisdom1980}. The value $\rI$ is larger for discs with higher initial-excitation levels, and kinks occur when $\rI$ is close to the nominal location of a strong MMR; this is especially true for discs with low initial eccentricities. Note that ${\rI/\aP}$ is independent of the planet's semimajor axis, and so depends only on the planet-to-star mass ratio and the disc's initial-excitation level.

\begin{figure}
	\includegraphics[width=8cm]{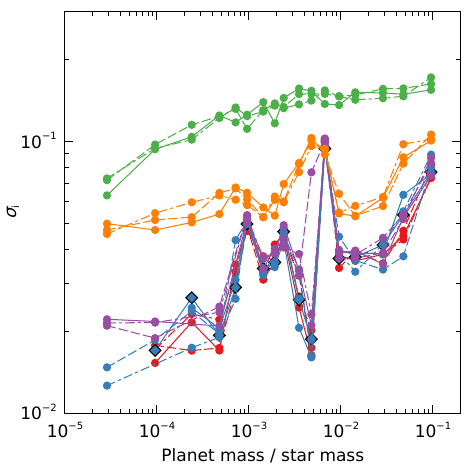}
    \caption{Flatness of the sculpted disc's inner edge, $\sigmaRI$, from fits to surface-density profiles at the end of our $n$-body simulations. A smaller value of $\sigmaRI$ corresponds to a sharper edge. Lines and symbols have the same meanings as on Figure \ref{fig: rIn_ap}. The edge steepness only depends on the planet-to-star mass ratio and the disc's initial-excitation level, and spikes in the plotted relationships occur when the inner edge is near the nominal location of a strong MMR.}
    \label{fig: sigmaRIn}
\end{figure}

Figure \ref{fig: sigmaRIn} shows how the flatness of the inner-edge profile, $\sigmaRI$, scales with the planet-to-star mass ratio and the initial disc-excitation level. It is independent of all other parameters. The inner edges are generally steeper if sculpted by lower-mass planets, and flatter for higher-mass planets. The inner edges are also flattest for discs with the highest initial-excitation levels, although there is less of a dependence for discs with low initial-excitation levels. The relationships between $\sigmaRI$ and $\mPOverMStar$ are also not smooth, but show complicated spikes. These spikes are not numerical effects, because they are replicated across simulations with the same mass ratios but different planet semimajor axes and star masses; the spikes actually occur when the inner edge is near the nominal location of a strong MMR.

The values of $\rI$ and $\sigmaRI$ on Figures \ref{fig: rIn_ap} and \ref{fig: sigmaRIn} were fitted for discs with initial surface-density profiles of $r^{-\alpha}$, where ${\alphaR=1.5}$, but they are actually independent of $\alphaR$ for realistic setups. This is because the planet-induced edges are typically much steeper than the overall disc profiles. To check this, we re-scaled the simulated discs to have initially flat surface-density profiles (i.e. ${\alpha=0}$) and re-fitted the edges. This resulted in values of $\rI$ and $\sigmaRI$ that are within a few percent of the ${\alpha=1.5}$ values, and hence the profiles of planet-sculpted edges do not strongly depend on the broader disc profiles. We further checked this analytically, and show that planet-sculpted inner edges should be much steeper than the overall disc provided that the initial-disc profile is flatter than ${r^{-4}}$ (and also ${r^{4}}$); this calculation is presented in \mbox{Appendix \ref{app: checkEdgeSteeperThanDisc}}. Most measured debris-disc profiles are shallower than this, so the steepness of a planet-sculpted edge should typically not depend on the overall disc profile.

\subsubsection{Edge-sculpting timescale}
\label{subsec: simulationsSDTimeEvolution}

The timescale for a circular-orbit planet to sculpt the disc inner edge is expected to scale with the diffusion timescale (Equations \ref{eq: diffusionTime} and \ref{eq: diffusionTimeUnits}), which quantifies how quickly a planet scatters and ejects debris. We find that this is reflected in our simulations; after some number of diffusion times, the inner edges settle into their final configurations. However, the number of diffusion timescales required appears to have some dependence on the planet-to-star mass ratio, with different regimes for ratios above and below $10^{-2}$.

For planet-to-star mass ratios below $10^{-2}$, we find that the inner edge settles into its final shape within 1 diffusion timescale; therefore, for the majority of realistic planets, Equations \ref{eq: diffusionTime} and \ref{eq: diffusionTimeUnits} are reasonable estimates of the sculpting timescale. However, for mass ratios above $10^{-2}$, more diffusion timescales are needed; such planets appear to take closer to 10 diffusion times to sculpt the disc. For this reason, we ran all simulations that had ${\mPOverMStar \geq 10^{-2}}$ until a time of ${100\tDiff}$, whilst all simulations with ${\mPOverMStar < 10^{-2}}$ were run for ${10\tDiff}$. Regardless, despite the difference in the number of diffusion times required, large planets still sculpt discs much faster than small planets, because the diffusion timescale strongly decreases with increasing planet mass.

There may be two reasons for this behaviour change around ${\mPOverMStar = 10^{-2}}$. First, it may mark the transition between a star-planet interaction, where ${\mPOverMStar \ll 1}$, to a binary interaction, where the two bodies have comparable mass. In the latter case, dynamical definitions like the Hill radius break down, so the interaction dynamics may fundamentally change.

The second reason relates to the ratio of $\tDiff$ to the planet's orbital period. For ${\mPOverMStar = 10^{-2}}$ the diffusion time is approximately 100 planet periods, and for ${\mPOverMStar = 10^{-1}}$ the diffusion time is just 1 planet period. For such mass ratios the approximations used to define the diffusion timescale start to break down. For example, a planet with ${\mPOverMStar = 10^{-1}}$ must take more than one period (and hence more than one $\tDiff$) to clear unstable material, because it can only eject material that passes close to it; any material that is nearly co-orbital with the planet but located on the opposite side of the star must therefore take several orbital periods (and hence several diffusion times) to pass close to the planet and get ejected. Conversely, for a small mass ratio the diffusion time is much longer than the planet period, so such effects are less important.

\subsection{Debris eccentricity at the disc inner edge}
\label{subsec: simulationsSDEccentricityOfInnerEdge}

To understand the profiles of the simulated inner edges, we must understand the dynamical processes occurring there. Like \cite{Marino2021} and \cite{Rafikov2023}, we expect the edge steepness to be related to debris eccentricity, so understanding the planet-induced eccentricity is vital for understanding how planets shape disc inner edges. In this section we quantify the eccentricity excitation at the simulated inner edges (Section \ref{subsec: simulationsMeasuringEccentricity}), and show that this excitation is caused by MMR interactions (Section \ref{subsec: simulationsMMRsCauseEccentricity}). Later, in Section \ref{sec: simpleAnalyticModel}, we will use these results to produce a predictive model relating planet properties to inner-edge profiles.

\subsubsection{Measuring debris eccentricity at the simulated inner edges}
\label{subsec: simulationsMeasuringEccentricity}

We first directly measure the eccentricities of inner-edge debris at the final snapshot of each of our $n$-body simulations. To do this, we fit the erf-powerlaw surface-density model (Equation \ref{eq: fittedProfile}) to each simulated disc, then use this to define the inner-edge region; the inner edge has a characteristic width of several times ${\rI \sigmaRI}$, so we define the `inner-edge region' as that centred on $\rI$ with an arbitrary full width ${3\rI \sigmaRI}$. We then calculate the rms eccentricity of all debris bodies that are instantaneously located in this radial range, which we define as the rms eccentricity of the debris-disc inner edge, $\eRMSI$. As an example, for this definition the simulation on \mbox{Figure \ref{fig: lowESurfaceDensities}} has an inner-edge region spanning from 12.5 to ${13.9\au}$, with ${\eRMSI=0.0433}$.

\begin{figure}
	\includegraphics[width=8cm]{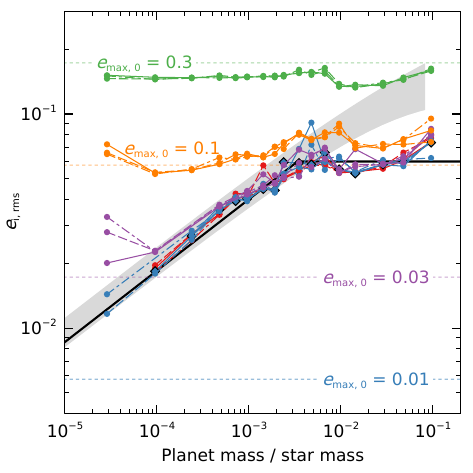}
    \caption{Root-mean-square eccentricities of surviving debris at the disc inner edges, at the end of our $n$-body simulations. Lines and symbols are defined on Figure \ref{fig: rIn_ap}. Dotted horizontal lines are the values expected if the discs maintain their initial-excitation levels; these should be compared to simulation lines of the matching colour. The shaded region is a theoretical prediction for the rms eccentricity acquired by bodies in the planet's 3:2 or 2:1 MMRs; its width is the spread of these values. The planet excites surviving debris through MMRs; the black line is a rough fit to the eccentricity it imparts on debris at the inner edge (Equation \ref{eq: planetERMSFit}). If this planet-induced eccentricity is smaller than the initial-disc excitation level, then the planet does not significantly increase the excitation level above that of the pre-interaction disc.}
    \label{fig: eRMSI}
\end{figure}

Figure \ref{fig: eRMSI} shows the rms eccentricities at the inner edges of each of our simulated discs, calculated using the above method. These rms eccentricities depend only on the planet-to-star mass ratio and the initial-disc excitation level. For discs with low initial-excitation levels, the eccentricity imparted by the planet increases with the planet-to-star mass ratio, up to mass ratios of ${\sim 3.3 \times 10^{-3}}$; above this, the inner-edge eccentricity is roughly independent of mass ratio. Conversely, if the disc's initial eccentricity exceeds the level that would be imparted on the edge by the planet, then the resulting edge eccentricity is essentially independent of the planet, and remains close to the initial level throughout the simulation.

\subsubsection{Mean-motion resonances as the edge-excitation mechanism for circular-orbit planets}
\label{subsec: simulationsMMRsCauseEccentricity}

We now identify the mechanism that excites inner-edge debris. Despite the significant eccentricities of this material (Figure \ref{fig: lowESurfaceDensities}, left panel), this excitation is not caused by planet-debris scattering, because the semimajor axes of this debris remains essentially unchanged. Excitation is also not due to secular interactions, because the low-eccentricity planet cannot sufficiently excite debris through secular interactions. This leaves MMRs as the only possible excitation mechanism, in agreement with previous studies \citep{Quillen2006Fom, QuillenFaber2006}. The idea that MMRs are responsible is also supported by the presence of debris populations with similar excited eccentricities near nominal MMR locations (e.g. the 2:1 MMR on Figure \ref{fig: lowESurfaceDensities}). Note that a single MMR does not usually dominate all surviving debris at the disc edge; rather, several nearby MMRs excite debris across a range of semimajor axes.

Given the resonant nature of debris at the inner edge, we can dynamically explain the dependence of ${\eRMSI}$ on planet-to-star mass ratio. The specific MMRs at the edge depend on the mass ratio, because as the mass ratio increases, the nominal MMR locations remain unchanged but the width of the chaotic zone around the planet increases. However, for low mass ratios the excitation level appears similar to that expected from the 2:1 or 3:2 MMRs, even if these are not the specific resonances at the edge. To show this, the grey region on Figure \ref{fig: eRMSI} is a theoretical prediction for the rms debris eccentricity expected for a population of bodies in the 2:1 or 3:2 MMRs, found by integrating the equations of motion for those MMRs (Pearce et al., in prep.). This is similar to  the ${(\mPOverMStar)^{1/3}}$ scaling from \cite{Petrovich2013} (their Equations 19 and 34), and agrees with our simulation results for lower planet-to-star mass ratios. Specifically, if a circular-orbit, non-migrating planet sculpts a debris-disc inner edge, then the rms eccentricity at that edge increases with planet mass, provided ${\mPOverMStar \lesssim 3.3 \times 10^{-3}}$.

For planet-to-star mass ratios above ${\sim 3.3 \times 10^{-3}}$, the inner-edge excitation is below that expected of 2:1 and 3:2 MMRs; this is because those MMRs would now lie inside the chaotic zone, and higher-order MMRs at semimajor axes outside the nominal 2:1 location are weaker and less effective at exciting debris (Figure \ref{fig: lowESurfaceDensities}). Hence for planet-to-star mass ratios above ${\sim 3.3 \times 10^{-3}}$, the debris eccentricity at the disc edge does not further increase with mass ratio.

Given this behaviour, we can roughly quantify the eccentricity that a planet on a circular orbit induces on initially unexcited debris at the disc inner edge:

\begin{equation}
  \eRMSIFromP \approx \begin{cases}
    0.4 \left(m_{\rm p} / m_*\right)^{1/3}, & \text{if $m_{\rm p} / m_* \leq 3.3 \times 10^{-3}$};\\
    0.06, & \text{else}.
  \end{cases}
  \label{eq: planetERMSFit}
\end{equation}

\noindent This is the black line on Figure \ref{fig: eRMSI}, which is a good match to the simulations where the planet-induced excitation dominates over the intrinsic disc excitation (e.g. the purple lines for mass ratios above $10^{-4}$). Conversely, since the initial-disc excitation would dominate over the planet-induced excitation if the former were high enough, we can generally predict the eccentricity of surviving debris at the inner edge of a planet-sculpted debris disc as

\begin{equation}
    \eRMSI = \max\left(\eRMSIFromP, \eRMSIZero\right).
    \label{eq: eRMSFromPlanetAndInitialDisc}
\end{equation}

\noindent where ${\eRMSIZero}$ is the `intrinsic' rms eccentricity at the disc edge (i.e. the pre-interaction eccentricity), and the planet-induced eccentricity $\eRMSIFromP$ is given by Equation \ref{eq: planetERMSFit}.

\section{Analytic model relating the inner-edge profile to the sculpting planet}
\label{sec: simpleAnalyticModel}

In Section \ref{sec: simulations} we showed that, if a circular-orbit planet sculpts a collisionless debris disc, then the steepness and relative location of the disc's inner edge depend only on the planet-to-star mass ratio and the disc's initial-excitation level. We also quantified how the planet excites debris eccentricities at the inner edge. Using these results, we now produce a simple analytical model to infer the properties of sculpting planets from inner-edge profiles. Section \ref{subsec: modelDesciption} details our analytical model, and Section \ref{subsec: modelExampleFit} demonstrates how to use it to infer the properties of an unseen sculpting planet from an observed inner-edge profile.

\subsection{Model setup and predictions}
\label{subsec: modelDesciption}

\begin{figure}
    \includegraphics[width=7cm]{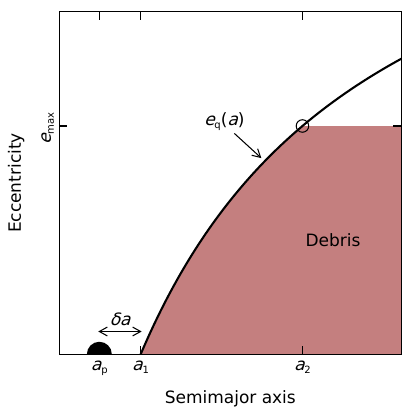}
    \caption{Schematic of our simple model for predicting the disc's inner-edge profile (Section \ref{sec: simpleAnalyticModel}). The black circle is the planet, and the red region surviving debris. Debris has eccentricities between 0 and ${e_{\rm max}}$. The planet eventually ejects all debris originating within its chaotic zone, which is any debris above the solid line (Equations \ref{eq: eqFromPeriAtOuterEdgeOfChaoticZone} and \ref{eq: deltaA}). The value ${a_1}$ is the smallest semimajor axis of any surviving particle (a circular orbit just exterior to the chaotic zone), and ${a_2}$ is the largest semimajor axis of any unstable particle (an orbit with the highest eccentricity ${e_{\rm max}}$, whose pericentre is at the outer edge of the chaotic zone).}
    \label{fig: scatteringModel}
\end{figure}

We assume a simplified model of the inner-edge region, as shown on Figure \ref{fig: scatteringModel}. We consider a planet on a circular orbit, and a debris disc that initially extends from the planet's semimajor axis out to some larger distance. The debris particles have initial eccentricities uniformly distributed between 0 and some $e_{\rm max}$; for this model, the origin of these eccentricities is unimportant. To predict the location and steepness of such a planet-sculpted disc edge, we consider which particles would be scattered by the planet, and which would survive.

In our model, any particle whose orbit comes within the planet's chaotic zone would be scattered and eventually  ejected. This means that, at late times, no particles should occupy the parameter space above the solid line on Figure \ref{fig: scatteringModel}; this line is the eccentricity ${e_{\rm q}(a)}$ that results in a particle's pericentre coinciding with the outer edge of the chaotic zone, defined by

\begin{equation}
    e_{\rm q}(a) \equiv 1 - \frac{\aP + \deltaA}{a}.
    \label{eq: eqFromPeriAtOuterEdgeOfChaoticZone}
\end{equation}

\noindent Here $a$ is the semimajor axis, and $\deltaA$ is the half-width of the chaotic zone, taken as 3 Hill radii for a circular-planet orbit:

\begin{equation}
    \deltaA \equiv 3 \rHill = 3 \aP \left(\frac{\mP}{3 m_*}\right)^{1/3}.
    \label{eq: deltaA}
\end{equation}

Since the planet eccentricity is zero, any non-scattered debris would retain its initial semimajor axis and eccentricity. Hence any debris with initial eccentricity above the Equation \ref{eq: eqFromPeriAtOuterEdgeOfChaoticZone} line would eventually get ejected, whilst any with eccentricity below the line would remain unperturbed. So in our model, at late times debris only occupies the shaded region of Figure \ref{fig: scatteringModel}.

In the following sections, we use this model to infer the location and steepness of the planet-sculpted inner edge. Before doing so, we must first define two final parameters: the semimajor axes $a_1$ and $a_2$, as shown on Figure \ref{fig: scatteringModel}. The span $a_1$ to $a_2$ roughly defines the inner-edge region in semimajor-axis space. The values of $a_1$ and $a_2$ correspond to the semimajor axes of orbits at the edge of the chaotic zone; $a_1$ corresponds to a circular orbit, and $a_2$ to an orbit with eccentricity equal to the maximum debris eccentricity $e_{\rm max}$. Expressions for $a_1$ and $a_2$ are given in \mbox{Appendix \ref{subsec: appA1AndA2}}.

\subsubsection{Predicted radial location of the disc inner edge, $\rI$}
\label{subsec: modelInnerEdgeLocation}

We now use the simple model on Figure \ref{fig: scatteringModel} to predict $\rI$, the characteristic location of a planet-sculpted inner edge. A full derivation of the $\rI$ prediction is presented in Appendix \ref{subsec: appRIDerivation}; our basic method is to first calculate the time-averaged radial location of all debris at a single semimajor axis $a$, and then average these for all debris with ${a_1 \leq a \leq a_2}$. This yields our prediction:

\begin{equation}
    \rI \approx \aP \left[1 + 3^{2/3} \left(\frac{\mP}{m_*}\right)^{1/3} \right] \left(1+\frac{2}{\sqrt{3}} \eRMSI \right),
    \label{eq: rInAnalyticEstimate}
\end{equation}


\noindent where we used ${\eRMSI = e_{\rm max}/\sqrt{3}}$ for a uniform $e$ distribution to convert $e_{\rm max}$ into $\eRMSI$.

Equation \ref{eq: rInAnalyticEstimate} predicts that the edge location depends on planet semimajor axis, planet mass, and debris eccentricity. The origin of this eccentricity is unimportant, provided that the eccentricity-semimajor axis distribution resembles that on Figure \ref{fig: scatteringModel}. We can therefore evaluate the eccentricity term in Equation \ref{eq: rInAnalyticEstimate} using Equations \ref{eq: planetERMSFit} and \ref{eq: eRMSFromPlanetAndInitialDisc}. If the debris eccentricity is set by the intrinsic (i.e. pre-interaction) eccentricity of the disc, which would occur if the intrinsic eccentricity is higher than that induced by the planet, then $\eRMSI$ in \mbox{Equation \ref{eq: rInAnalyticEstimate}} can simply be replaced by the intrinsic disc eccentricity $\eRMSIZero$. Alternatively, for a disc with sufficiently low pre-interaction eccentricity, we can substitute ${\eRMSI=\eRMSIFromP}$ from Equation \ref{eq: planetERMSFit} to predict the location of the planet-sculpted inner edge:

\begin{equation}
  \rI \approx \begin{cases}
    \aP \left[1 + 2.5 \left(\frac{\mP}{m_*}\right)^{1/3} \right], & \text{if $m_{\rm p} / m_* \leq 3.3 \times 10^{-3}$}; \vspace{2mm} \\
    \aP \left[1.1 + 2.2\left(\frac{\mP}{m_*}\right)^{1/3} \right], & \text{else},
  \end{cases}
  \label{eq: rInAnalyticEstimatePlanetDrivenE}
\end{equation}

\noindent where we omit higher-order terms in $\left(\mPOverMStar \right)^{1/3}$.

\begin{figure}
	\includegraphics[width=8cm]{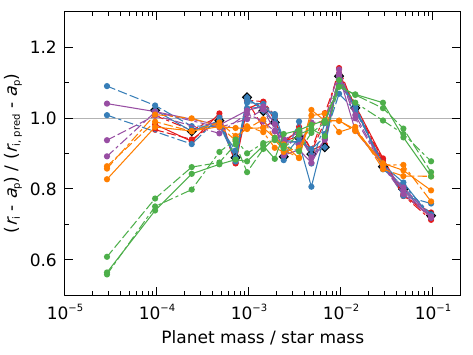}
    \caption{Comparison of our predicted inner-edge location to the results of $n$-body simulations. The vertical axis is the difference between the planet orbit and the inner-edge location $\rI$ fitted from the simulations, divided by the equivalent value predicted by \mbox{Equation \ref{eq: rInAnalyticEstimate}}. The horizontal grey line denotes unity; all other symbols and lines are defined on Figure \ref{fig: rIn_ap}. Our predicted edge locations are generally similar to those from the simulations, with the predictions typically within ${10\percent}$ of the simulated values for planets below ${1\percent}$ of the stellar mass and initial debris-excitation levels of $\eMaxZero < 0.3$. Predictions are worse for larger mass ratios, and also for simulations with both high initial-debris eccentricities and low-mass planets.}
    \label{fig: comparingRInPredToSims}
\end{figure}

Having made a prediction for $\rI$, we now compare this prediction to the results of our $n$-body simulations. For each simulation, we predict $\rI$ by evaluating Equation \ref{eq: rInAnalyticEstimate}; to do this, we take $\aP$ and $\mP$ from the simulation setup, and also use the value of $\eRMSI$ measured directly from the simulation. The results are shown on Figure \ref{fig: comparingRInPredToSims}. We see that the prediction generally works well; the predicted value of ${\rI - \aP}$ for simulations with $\eMaxZero < 0.3$ is typically within ${10\percent}$ of that from simulations. However, the prediction is less accurate for discs with very high intrinsic eccentricities ($\eMaxZero = 0.3$) that are interacting with low-mass planets (${\mPOverMStar \lesssim 10^{-4}}$); for such high eccentricities there is considerable overlap of debris orbits, so our simple $\rI$ approximation using only the average debris position probably no longer holds. There is also a noticeable divergence for planet-to-star mass ratios above ${10^{-2}}$, which could be due to a fundamental shift in the dynamics; this is discussed in Section \ref{subsec: simulationsSDTimeEvolution}. Nonetheless, \mbox{Equation \ref{eq: rInAnalyticEstimate}} holds across the large majority of our explored parameter space, so offers a reasonable means to predict the location of a planet-sculpted inner edge.

\subsubsection{Predicted steepness of the disc inner edge, $\sigmaRI$}
\label{subsec: modelInnerEdgeSteepness}

In this section we use our simple model to predict the shape of the planet-sculpted inner edge, as quantified by $\sigmaRI$. Near the inner edge, our fitted profile (Equation \ref{eq: fittedProfile}) is essentially an erf function; this has characteristic width ${x \rI \sigmaRI}$, where $x$ is a scalar of order unity. We hypothesise that this width should be roughly equivalent to the distance between the pericentre of the innermost stable particle and the apocentre of the outermost unstable particle, i.e. ${x \rI \sigmaRI \approx a_2(1+e_{\rm max}) - a_1}$ from Figure \ref{fig: scatteringModel}. Substituting expressions for $\rI$, $a_1$ and $a_2$ yields ${\sigmaRI \propto e_{\rm max}}$ to first order in $e_{\rm max}$, i.e. the steepness of a planet-sculpted inner edge depends entirely on the eccentricity of debris around that edge. \cite{Marino2021} and \cite{Rafikov2023} reached a similar conclusion for edges with sharp cutoffs in semimajor axis, but we show that this proportionality is also expected from the smooth distribution arising from scattering.

\begin{figure}
	\includegraphics[width=8cm]{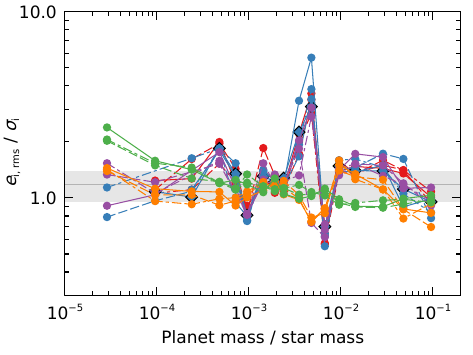}
    \caption{Relating the steepness of sculpted-disc inner edges to the debris eccentricities at those edges, from our $n$-body simulations. Smaller $\sigmaRI$ values correspond to steeper edges. Our simple model predicts $\sigmaRI$ and $\eRMSI$ to be proportional, in agreement with the simulations. The horizontal grey line and shaded region show the value and uncertainty of the prefactor ${1.2 \pm 0.2}$ in Equation \ref{eq: rSigInAnalyticEstimate}, which is empirically measured from the simulations. All other lines and symbols are defined on Figure \ref{fig: rIn_ap}. The breakdown in the fit at planet-to-star mass ratios between $10^{-3}$ and $10^{-2}$ for low-eccentricity simulations is due to the 2:1 MMR imposing additional structure on the edge (discussed in Section \ref{subsec: simulationsSDFitResultsAtLateTimes}).}
    \label{fig: eRMSI_rSigIn}
\end{figure}

Since we expect ${\sigmaRI \propto e_{\rm max}}$, we use our $n$-body simulations to directly relate $\sigmaRI$ to debris eccentricity. Figure \ref{fig: eRMSI_rSigIn} shows the rms eccentricities measured at the inner edges of our simulated discs, divided by the $\sigmaRI$ values fitted to those simulations. This ratio is roughly constant for all simulations, verifying our prediction that $\sigmaRI$ depends only on $e_{\rm max}$ (and hence $\eRMSI$). Taking the median of this ratio from our simulations, we find that the flatness of the inner edge of a planet-sculpted disc depends on the eccentricity of debris at that edge as

\begin{equation}
    \eRMSI \approx (1.2 \pm 0.2) \; \sigmaRI.
    \label{eq: rSigInAnalyticEstimate}
\end{equation}

\noindent Here the uncertainty on the empirical prefactor is defined from the inter-quartile range on ${\eRMSI / \sigmaRI}$ from our simulations.

\subsection{Inferring the properties of a sculpting planet from the location and shape of a debris-disc inner edge}
\label{subsec: modelExampleFit}

Section \ref{subsec: modelDesciption} related the disc inner edge to the properties of a sculpting planet. Since it has historically been easier to resolve a cold debris disc than to detect a distant planet, it is common to use observed debris discs to infer the properties of unseen planets (e.g. \citealt{Pearce2022ISPY}). In this section we provide a method to constrain a sculpting planet on a circular orbit from the shape and location of a debris-disc inner edge.

\begin{figure}
	\includegraphics[width=7cm]{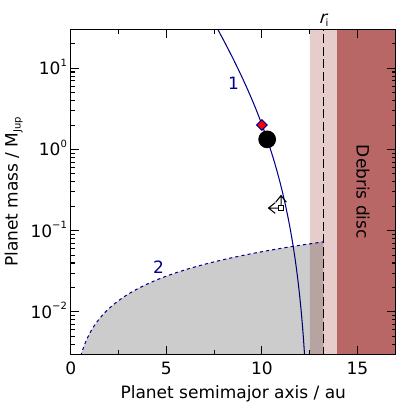}
    \caption{Example application of our results: using a debris disc's inner-edge profile to infer the properties of an unseen sculpting planet. The plot shows the disc from the $n$-body simulation on \mbox{Figure \ref{fig: lowESurfaceDensities}}, for which we attempt to constrain the planet using the step-by-step method from Section \ref{subsec: modelExampleFitStepByStepMethod}. The disc has a fitted inner-edge radius ${\rI=13.2\au}$ and flatness ${\sigmaRI=0.0361}$. The dashed vertical line is $\rI$, and the light brown region around it is the approximate full width of the inner edge ($3\rI\sigmaRI$). The large black circle shows the planet properties we infer, and the red diamond is the actual planet in the $n$-body simulation. The white square is the planet's minimum-possible mass and maximum semimajor axis found using the alternative method of \citet{Pearce2022ISPY}, which accounts for the edge location but not its steepness. Line 1 shows the planet properties required to set the edge at $\rI$ (Equation \ref{eq: mPltFromRI}), and Line 2 is the minimum planet mass required to have sculpted the disc edge within the age of the star (Equation \ref{eq: minMPMassFromDiffusionTime}); for this example the age is arbitrarily taken as ${100\myr}$.}
    \label{fig: examplePredictedPlanet}
\end{figure}

To demonstrate the method, we apply it to the simulation on \mbox{Figure \ref{fig: lowESurfaceDensities}} as a example. We assume that the simulated disc has been observed, but that the sculpting planet is undetectable. We will constrain the properties of the unseen planet from the disc alone, then compare these to the known parameters of the simulated planet to gauge the effectiveness of the method. The various steps in the calculation are shown on \mbox{Figure \ref{fig: examplePredictedPlanet}}, and described below. For this example we assume that the planet has finished sculpting the disc by the time the observations are made.

\subsubsection{Step-by-step method}
\label{subsec: modelExampleFitStepByStepMethod}

The first step is to fit the disc surface-density profile with some function that quantifies the location and steepness of the inner edge. In this paper we use an erf-powerlaw function (Equation \ref{eq: fittedProfile}), which yields inner-edge location $\rI$ and flatness $\sigmaRI$; other parameterisations could alternatively be used, and in Appendix \ref{app: relatingDifferenentEdgeProfileModels} we provide equations to convert the outputs of common parametric models into $\rI$ and $\sigmaRI$ values. For our example simulation, the fitted inner-edge profile has ${\rI = 13.2\au}$ and ${\sigmaRI=0.0361}$. 

The second step, now that we have the edge profile, is to infer the sculpting-planet mass as a function of its semimajor axis. This step makes the implicit assumption that the planet has a circular orbit and has not migrated, but it otherwise applies regardless of whether the planet or some other process is responsible for debris eccentricities. \mbox{Equation \ref{eq: rInAnalyticEstimate}} relates planet mass and semimajor axis to the location and rms eccentricity of the disc inner edge; using this equation, and noting the relation between $\eRMSI$ and $\sigmaRI$ (Equation \ref{eq: rSigInAnalyticEstimate}), we can infer the sculpting-planet mass using


\begin{equation}
    \mP \approx 116 \mJup \left(\frac{m_*}{\mSun}\right) \left[\frac{\rI}{\aP} \left(1 + 1.35 \sigmaRI \right)^{-1} -1 \right]^3.
    \label{eq: mPltFromRI}
\end{equation}

\noindent This is Line 1 on Figure \ref{fig: examplePredictedPlanet} (noting that ${m_* = 1\mSun}$ in our example).

The next step is to put a lower bound on the planet mass, assuming that the planet has finished sculpting the observed-disc edge. This means that the sculpting timescale must be shorter than the star \mbox{age $t_*$}. In Section \ref{subsec: simulationsSDTimeEvolution} we showed that the sculpting timescale is approximately the diffusion time if ${\mPOverMStar \lesssim 10^{-2}}$, and ${\sim 10}$ diffusion times otherwise; we can therefore rearrange Equation \ref{eq: diffusionTimeUnits} and evaluate it at ${a=\rI}$ to show that 

\begin{equation}
    \mP \gtrsim 0.105 \sqrt{k} \mJup \left(\frac{t_*}{\myr}\right)^{-1/2} \left(\frac{\aP}{\au}\right) \left(\frac{m_*}{\mSun}\right)^{3/4}\left(\frac{\rI}{\au}\right)^{-1/4} ,
    \label{eq: minMPMassFromDiffusionTime}
\end{equation}

\noindent where

\begin{equation}
  k \approx \begin{cases}
    1, & \text{if $\mPOverMStar \lesssim 10^{-2}$};\\
    10, & \text{else}.
  \end{cases}
  \label{eq: numberOfDiffusionTimes}
\end{equation}

\noindent For our example we ascribe an arbitrary stellar age of ${100\myr}$, which results in \mbox{Line 2} on \mbox{Figure \ref{fig: examplePredictedPlanet}}.

The final step is to use the edge shape to break the degeneracy between planet mass and semimajor axis. Unlike previous steps, this step requires the implicit assumption that the planet is solely responsible for exciting debris. It is also only valid if ${\sigmaRI \lesssim 0.05}$ \mbox{(i.e. ${\eRMSI \lesssim 0.06}$)}; this is the flattest profile that a non-migrating, circular-orbit planet could impart on an initially low-eccentricity disc (Figure \ref{fig: eRMSI}), so if ${\sigmaRI \gtrsim 0.05}$ then some other process must have excited debris and this step cannot be applied. If ${\sigmaRI \lesssim 0.05}$, then rearranging \mbox{Equation \ref{eq: planetERMSFit}} yields the planet mass as

\begin{equation}
    \mP \approx 2.64\times 10^4 \mJup \left(\frac{m_*}{\mSun}\right) \sigmaRI^3 ;
    \label{eq: planetMassFromSigma}
\end{equation}

\noindent we can then substitute this mass into Equation \ref{eq: mPltFromRI} to yield the planet's semimajor axis as

\begin{equation}
    \aP \approx \rI \left(1+1.35\sigmaRI\right)^{-1} \left[ 1 + 0.205 \left(\frac{\mP}{\mJup}\right)^{1/3} \left(\frac{\m_*}{\mSun}\right)^{-1/3} \right]^{-1}.
    \label{eq: planetSemimajorAxisFromSigma}
\end{equation}


\noindent For our example, this yields a planet mass of ${1.24\mJup}$ and semimajor axis ${10.3\au}$; these are in good agreement with the actual values of ${2\mJup}$ and ${10\au}$ from the simulation, as shown on Figure \ref{fig: examplePredictedPlanet}. Note that caution must be applied if ${\sigmaRI \sim 0.05}$ \mbox{(i.e. ${\eRMSI \sim 0.06}$)}, because such an edge profile could be imparted by any planet with ${\mPOverMStar \gtrsim 3.3 \times 10^{-3}}$ (Figure \ref{fig: eRMSI}); in this case, the mass estimate from Equation \ref{eq: planetMassFromSigma} should be interpreted as a lower bound rather than a single value.

\subsubsection{Accuracy of planet parameters inferred from disc edges}
\label{subsec: modelAccuracyOfInferredPlanets}

\begin{figure}
	\includegraphics[width=8cm]{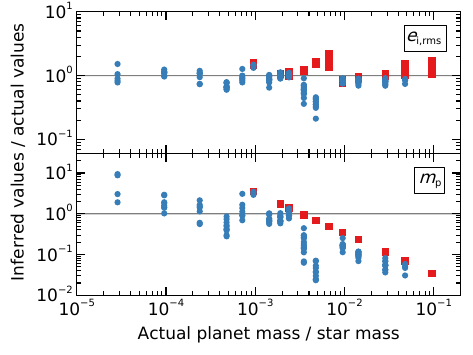}
    \caption{Planet properties inferred from simulated-disc inner edges using the process in Section \ref{subsec: modelExampleFit}, compared to the actual planets. Top panel: ${\eRMSI}$ values inferred from our fitted inner-edge profiles (Equation \ref{eq: rSigInAnalyticEstimate}), which agree well with the actual values. Bottom panel: planet masses inferred from the edge steepnesses (Equation \ref{eq: planetMassFromSigma}), which agree well with the actual values for planet-to-star mass ratios below ${3.3 \times 10^{-3}}$ (equivalent to ${3.5\mJup}$ around a Sun-like star). For higher ratios, Equation \ref{eq: planetMassFromSigma} can significantly underpredict the planet mass, because the edge excitation level becomes constant with planet mass (Figure \ref{fig: eRMSI}). Note that this degeneracy only affects planet masses that were directly inferred from the edge steepness (black circle on Figure \ref{fig: examplePredictedPlanet}); regardless of this, all planets would still lie along the mass-semimajor axis relation (Line 1 on Figure \ref{fig: examplePredictedPlanet}). Blue circles and red squares are simulations where the inferred debris excitation ${\eRMSI}$ at the disc inner edge is below and above 0.06, respectively (where ${\eRMSI \approx 0.06}$ corresponds to an edge profile with ${\sigmaRI \approx 0.05}$).}
    \label{fig: comparisonToActual}
\end{figure}

The above example showed that planet parameters can be well inferred from inner-edge profiles in at least some setups. We now repeat the above process for a large fraction of our simulations, to test how well the method applies in general. We omit simulations where the initial disc-excitation level is larger than the expected excitation generated by the planet, because for those simulations, Equation \ref{eq: planetMassFromSigma} cannot be used to infer planet mass. The results are shown on \mbox{Figure \ref{fig: comparisonToActual}}. The top panel shows that the inferred debris-excitation level $\eRMSI$, calculated from the edge profile using Equation \ref{eq: rSigInAnalyticEstimate}, agrees with the actual simulation values in almost all cases; the largest discrepancy is for planet-to-star mass ratios around ${5 \times 10^{-3}}$, which is where additional structure from the strong \mbox{2:1 MMR} coincides with the disc edge. The bottom panel shows that the planet masses inferred using Equation \ref{eq: planetMassFromSigma} agree with the actual planet masses for low mass ratios, but that the two diverge if the actual planet has ${\mPOverMStar \gtrsim 3.3 \times 10^{-3}}$. For the highest mass ratios, \mbox{Equation \ref{eq: planetMassFromSigma}} can underpredict the planet mass by over one order of magnitude. The reason for this is that, for planets with mass ratios above ${3.3 \times 10^{-3}}$, the level of eccentricity excitation imparted on the disc edge is independent of planet mass (Figure \ref{fig: eRMSI}); this can cause Equation \ref{eq: planetMassFromSigma} to significantly underpredict the planet mass if debris at the disc edge has an excitation level of ${\eRMSI\approx0.06}$. Note that this degeneracy only affects the use of Equation \ref{eq: planetMassFromSigma}; regardless of the mass ratio, and whether or not the planet was responsible for exciting debris, a sculpting planet should still lie close to the mass-semimajor axis relation from Equation \ref{eq: mPltFromRI} (Line 1 on Figure \ref{fig: examplePredictedPlanet}).

\section{Collisions}
\label{sec: collisions}

Sections \ref{sec: simulations} and \ref{sec: simpleAnalyticModel} assessed how a circular-orbit planet would sculpt the inner edge of a debris disc, if the interaction were purely $n$-body dynamics. For that scenario we showed that the edge steepness could be directly related to the planet mass. However, in a real disc there are also collisions between debris bodies, which would alter the edge profile. In this section we consider the collisional evolution, and answer two basic questions: could a collision-dominated disc edge be differentiated from a planet-dominated edge, and how would collisions change the profile of a planet-sculpted edge?

\subsection{Could a purely collisional disc be differentiated from a purely planet-sculpted disc?}
\label{subsec: collisionsCanPlanetAndCollisionsBeDifferentiated}

Destructive collisions cause debris discs to lose mass over time; material grinds down to dust, which is eventually removed by radiation forces \citep{Wyatt1999,Wyatt2008,Krivov2010}. The rate at which this process occurs depends on several main factors: the number density of debris bodies, their size distribution, and their relative velocities. 

For an initially broad debris disc, these factors result in well-defined collisional evolution. At early times the surface density is expected to decrease with distance, for example going as ${r^{-1.5}}$ in the MMSN. Relative velocities are also higher in the inner regions, and this combination makes the initial collisional depletion faster closer to the star. However, this means that the surface density closer to the star drops faster than that further out, reducing the inner collision rate. The result is that the initially negative surface-density slope gradually transforms into a positive one; a `collision front' moves outwards across the disc, where the region interior to the front tends towards a positive surface-density slope, whilst the exterior region still has its original profile. In the collisionally processed region, this profile is well-characterised as a powerlaw $r^\pI$, with index ${\pI = 7/3 = 2.3}$ in \cite{Wyatt2007} and \cite{Kennedy2010}, or ${\pI=2}$ in more-recent models \citep{Lohne2008, ImazBlanco2023}.

Conversely, if a planet sculpts a collisionless disc, then the inner-edge steepness is set by the planet mass. To answer whether such a planet-dominated disc can be differentiated from a collision-dominated disc, we must compare the inner-edge slopes. Using Equation \ref{eq: appSigmaRIEqualsPowerlawSteepness} to relate an erf-fitted slope to a powerlaw slope, we see that a planet-dominated disc will have a steeper inner edge than a collision-dominated disc if ${\sigmaRI < 0.40}$ (assuming ${\pI=2}$ for the collisional disc; if ${\pI = 7/3}$ were used instead, then ${\sigmaRI < 0.35}$ for the sculpted edge to be steeper). Figure \ref{fig: sigmaRIn} shows that these conditions are satisfied by the end of all of our circular-planet simulations, meaning that the inner edge of a purely planet-sculpted disc is steeper than that of a purely collisional disc if sculpting has completed. In some cases it would be much steeper; our steepest $n$-body discs have ${\sigmaRI \sim 10^{-2}}$, which would correspond to ${\pI > 10}$ in $r^\pI$ notation. So we should be able to differentiate purely planet-sculpted discs from purely collision-dominated discs, with the former having much steeper inner edges.

\subsection{How would collisions affect a planet-sculpted edge?}
\label{subsec: collisionsHowWouldCollisionsAffectEdge}

The previous section compared a collisionless, planet-sculpted disc to a collisional disc without a planet. However, in reality a planet-sculpted disc would also undergo collisions. In this section we consider collisions in more detail, to assess how they would affect a planet-sculpted disc.

A fully self-consistent model, where a disc undergoes $n$-body interactions whilst simultaneously collisionally evolving, is difficult to properly implement. In particular, there are several parameters that are critical for quantifying the collisional evolution, but that are fundamentally unknown for real debris discs; these include the debris-disc mass, and the sizes of the largest planetesimals \citep{KrivovWyatt2021}. For this reason we chose to decouple the dynamical and collisional modelling in this paper. Instead, we will take the disc morphologies arising from our $n$-body simulations, and then input these into a collisional model to estimate how collisions would modify a planet-sculpted disc edge. This approach is not fully self-consistent, but would be valid if the planet-sculpting timescale were much shorter than the collisional timescale; we will later show that this condition holds for many plausible scenarios.

\subsubsection{Collisional model}
\label{subsec: collisionsLohneModel}

To model collisions, we use a {\sc mathematica} implementation of the \cite{Lohne2008} collisional model. The model takes an axisymmetric surface-density profile, and splits it into radial bins. Each bin is then treated separately, and the dust mass in each bin is determined after some time. We then use the masses in each radial bin to calculate the dust surface-density profile. In this section we briefly describe the model and the parameters we use; we refer the reader to \cite{Lohne2008} for a much more detailed description.

We assume the eccentricity $e$ and inclination $I$ of the debris population are related by $I \approx e/2$, and treat $e$ as a free parameter. Following \cite{Lohne2012}, \cite{Schuppler2016} and \cite{KrivovIda2018}, we assume a critical fragmentation energy of

\begin{equation}
    \qDStar = \left(\frac{v_{\rm col}}{v_0} \right)^{1/2}
    \left[A_{\rm s} \left(\frac{s}{1\m}\right)^{3\bS}
    + A_{\rm g} \left(\frac{s}{1\km}\right)^{3\bG} \right]
    \label{eq: QDStar}
\end{equation}

\noindent with ${v_0=3\kmPerS}$, ${A_{\rm s}=A_{\rm g}=5\times10^6 \; {\rm erg \; g^{-1}}}$, ${b_{\rm s}=-0.12}$ and ${b_{\rm g}=0.46}$. The planetesimal-collision speed $v_{\rm col}$ is given by

\begin{equation}
    v_{\rm col} = \sqrt{\frac{\mathcal{G} m_*}{r}} f(e, I),
    \label{eq: vCol}
\end{equation}

\noindent where $\mathcal{G}$ is the gravitational constant and

\begin{equation}
    f(e, I) = \sqrt{\frac{5}{4} e^2 + I^2}.
    \label{eq: fInCollisionModel}
\end{equation}

\noindent The value of $\qDStar$ is dominated by the material strength for smaller bodies, and gravity for larger bodies. The transition between these can be defined either as the size at which the strength and gravity terms in Equation \ref{eq: QDStar} are equal, or as the size at which $\qDStar$ is minimised. Following \cite{Lohne2008}, here we use the former definition. For our parameters, the strength-gravity transition then occurs at the ``breaking'' radius ${\sB=232\m}$.

We assume that debris in each bin follows a three-slope size distribution as in \cite{Lohne2008}, with slope ${\qP=1.87}$ for primordial bodies, ${\qG=1.68}$ in the gravity-dominated quasi-steady state and ${\qS=1.89}$ in the strength-dominated quasi-steady state. We choose minimum and maximum dust-grain radii of ${\sMin=2\um}$ and ${\sD = 1\mm}$, and treat the largest-planetesimal radius $\sMax$ as a free parameter. We use a bulk density of solids ${\rho=3.5\gPerCmCubed}$.

According to the model, there are two key timescales that determine the collisional evolution. The shorter, $\tauB$, is the time when the weakest bodies begin to collide. At this time, the collisional decay of the dust density sets in (if $\qP\le \qS$) or at least speeds up (if $\qP > \qS$); see Equation 43 and Figure 9 in \cite{Lohne2008}. The value of $\tauB$ is given by Equation 31 in \cite{Lohne2008}:

\begin{multline}
    \tauB = \frac{16 \pi \rho}{3 M_0} \left(\frac{\sB}{\sMax} \right)^{3\qP-5} \frac{\sMax r^{7/2}}{\sqrt{\mathcal{G} m_*}} \frac{\delta r}{r} \\
    \times \frac{\qP -5/3}{2-\qP} \left[ 1 - \left(\frac{\sMin}{\sMax} \right)^{6-3\qP}\right] \frac{I}{f(e,I) G(\qP, \sB, r)},
    \label{eq: tauB}
\end{multline}

\noindent where $M_0$ is the total initial mass of solids in the bin, ${\delta r}$ is the radial width of the bin,

\begin{multline}
    G(q, s, r) = \left[\xC(s, r)^{5-3q}-\left(\frac{\sMax}{s}\right)^{5-3q} \right]\\
    + 2\frac{q-5/3}{q-4/3}\left[\xC(s, r)^{4-3q}-\left(\frac{\sMax}{s}\right)^{4-3q} \right]\\
    + \frac{q-5/3}{q-1}\left[\xC(s, r)^{3-3q}-\left(\frac{\sMax}{s}\right)^{3-3q} \right],
    \label{eq: GInCollisionModel}
\end{multline}

\noindent

and

\begin{equation}
    \xC(s,r) = \left[\frac{2 \qDStar(s,r) r}{f^2(e, I) \mathcal{G} m_*}\right]^{1/3}.
    \label{eq: XcInCollisionModel}
\end{equation}

\noindent We arbitrarily use ${\delta r/r = 0.1}$ in our code, but this later cancels out so its value does not affect any results.

The longer timescale is $\tauMax$: the collisional lifetime of the largest planetesimals. This is Equation 42 in \cite{Lohne2008}\footnote{Equation 42 in \cite{Lohne2008} has an erroneous index of -1 around the square bracket, which we omit here; this does not affect our results because ${\sMin\ll\sMax}$.}:

\begin{multline}
    \tauMax = \frac{16 \pi \rho}{3 M_0} \frac{\sMax r^{7/2}}{\sqrt{\mathcal{G} m_*}} \frac{\delta r}{r} \\
    \times \frac{\qG -5/3}{2-\qP} \left[ 1 - \left(\frac{\sMin}{\sMax} \right)^{6-3\qP}\right] \frac{I}{f(e,I) G(\qG, \sMax, r)}.
    \label{eq: tauMax}
\end{multline}

\noindent We will later show that $\tauMax$ is the most important factor for determining the collisional behaviour. We provide an open-access {\sc python} code to calculate $\tauMax$ given the assumptions in this paper\footnote{http://www.tdpearce.uk/public-code}, and in Section \ref{subsec: collisionsImplications} we write $\tauMax$ in an alternative form to make its dependencies more explicit.

Given the above setup, the model yields the dust mass $m_{\rm dust}$ in each radial bin after some time $t$. This is Equation 43 in \cite{Lohne2008}:

\begin{multline}
    m_{\rm dust}(t) = \frac{M_0}{1+t/\tauMax} \mathcal{T}^{(\qG-\qP)/[\qP-5/3+(\qP-1)\bG]} \frac{2-\qP}{2-\qS} \\
    \times \left(\frac{\sB}{\sMax}\right)^{6-3\qP} \left[\left(\frac{\sD}{\sB}\right)^{6-3\qS} - \left(\frac{\sMin}{\sB}\right)^{6-3\qS} \right]^{-1}
    \label{eq: mDustAtT}
\end{multline}

\noindent for ${\tauB < t < \tauMax}$, where ${\mathcal{T}=t/\tauB}$ if ${t<\tauMax}$ or ${\tauMax/\tauB}$ otherwise\footnote{Equation 43 in \citet{Lohne2008} has erroneous indices of ${2-q}$ instead of ${6-3q}$, for $\qP$ and $\qS$.}. For ${t<\tauB}$, $\qG$ and $\bG$ should be replaced by $\qS$ and $\bS$ respectively.

For each collisional simulation, we initialise the disc with a surface-density profile similar to the outcomes of our $n$-body simulations. This is a powerlaw with a truncated inner edge:

\begin{equation}
    \SigmaR = \xM \left(\frac{r}{\au}\right)^{-\alphaR} \frac{\mEarth}{\au^2} \times \frac{1}{2} \left[ 1-{\rm erf}\left(\frac{\rI - r}{\sqrt{2} \sigmaRI \rI} \right) \right],
    \label{eq: collisionSimInitialSDProfile}
\end{equation}

\noindent where a disc with ${\xM=1}$ and ${\alphaR=3/2}$ would have the MMSN profile beyond the inner edge. This initial profile sets $M_0$ in the above equations. The initial values of ${\xM}$ and ${\alphaR}$ are free parameters that we vary between models. After setting up the disc, we use \mbox{Equation \ref{eq: mDustAtT}} to determine the dust mass in each radial bin after some time. Using these bin masses, we compute the dust surface-density profile. Finally, we fit this profile with an erf-powerlaw function similar to Equation \ref{eq: collisionSimInitialSDProfile}, to measure $\rI$, $\sigmaRI$ and $\alphaR$ at that time.

\subsubsection{Collision results}
\label{subsec: collisionsResults}

The previous section described the collisional model. In this section we apply the model to assess the impact of collisions on planet-sculpted disc edges.

\begin{figure}
	\includegraphics[width=7cm]{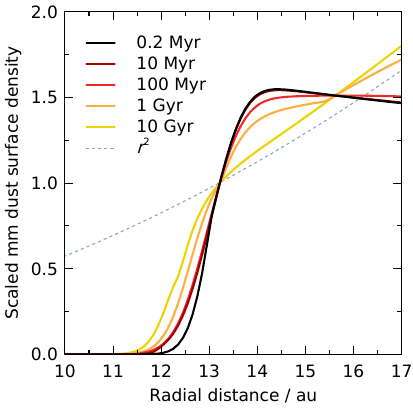}
    \caption{Collisional evolution of the planet-sculpted disc from Figure \ref{fig: lowESurfaceDensities}, according to the \citet{Lohne2008} model (Section \ref{subsec: collisionsHowWouldCollisionsAffectEdge}). We initialise the disc with ${\rI=13.2\au}$ and ${\sigmaRI=0.0361}$, and assume ${m_*=1\mSun}$, $\xM=1$ and $\sMax=50\km$. This yields ${\tauMax=1.1\gyr}$ and ${\tauB=0.14\myr}$ at ${13.2\au}$. Solid lines show the edge at various times; collisions flatten the edge, tending towards an $r^2$ profile (dashed line). Note that our simulated collisions only deplete material; the lines are scaled to the same value at ${13.2\au}$ for gradient comparison, giving the impression that the surface density increases with time at some radii, but in reality it decreases at each successive snapshot and unscaled lines would not cross (the lines are scaled because the surface density drops considerably with time, and a logarithmic plot would skew the gradient comparison). The edge undergoes little evolution until the time becomes comparable to $\tauMax$; after that, the edge starts to flatten.}
    \label{fig: collisionEvoSingleSetup}
\end{figure}

We first apply it to the $n$-body disc from Figure \ref{fig: lowESurfaceDensities} as an example. We initialise the disc with ${\xM=1}$ and ${\alphaR=3/2}$, with an inner edge defined by ${\rI=13.2\au}$ and ${\sigmaRI=0.0361}$, and set the largest-planetesimal radius to ${\sMax=50\km}$. We also set the debris eccentricity to 0.043, the rms eccentricity at the disc inner edge in the $n$-body simulation. This results in  ${\tauMax=1.1\gyr}$ and ${\tauB=0.14\myr}$ at ${13.2\au}$ (Equations \ref{eq: tauB} and \ref{eq: tauMax}). We then use the collisional model to calculate the edge profile at various times up to ${10\gyr}$, the lifetime of the solar-type star. The first snapshot is made shortly after $\tauB$, to ensure there is sufficient dust in the system. Figure \ref{fig: collisionEvoSingleSetup} shows the results; the inner edge is largely unchanged for the first ${100\myr}$, but from ${1\gyr}$ collisions begin to make the edge flatter. We quantify this flattening using the erf-powerlaw function; at 0.2, 10, 100, $10^3$ and ${10^4 \myr}$, the respective fitted values are  ${\rI = 13.1}$, 13.0, 13.0, 13.0 and ${13.2\au}$, and ${\sigmaRI = 0.034}$, 0.042, 0.043, 0.064 and 0.11. Note that the erf function is an increasingly poor fit to the edge shape as $\tauMax$ is approached, due to the complex edge profile.

Figure \ref{fig: collisionEvoSingleSetup} demonstrates the importance of $\tauMax$ in setting the collisional evolution. Before $\tauMax$, collisions have little effect on the inner-edge profile. However, once $\tauMax$ is reached, collisions start to have a significant effect, making the edge flatter. This collisional profile is expected to eventually tend towards $r^2$, the profile for a broad collisional disc, as shown by the dashed line on Figure \ref{fig: collisionEvoSingleSetup}. However, in this example the profile is never reached within the stellar lifetime; in fact, the inner edge assumes two slopes, with an erf-like profile interior to $\rI$ and a flatter, positive profile beyond this. This means that, in this example, the effect of planet sculpting would still be visible on the disc's inner-edge even at very late times\footnote{Since dust mass, and hence brightness, decreases with time, a real observation of a collisionally evolved disc might require a high signal-to-noise ratio to detect the deviation from an $r^2$ profile at $\rI$.}.

\begin{figure}
	\includegraphics[width=7cm]{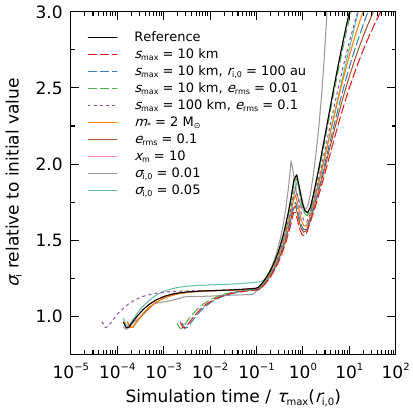}
    \caption{General collisional evolution of the inner edge of a planet-sculpted debris disc. The plot shows the evolution of the inner-edge profile over time, where time is expressed relative to $\tauMax(\rIZero)$, the collisional lifetime of the largest planetesimals at the initial inner edge. The simulations start at time $\tauB$. The black line is the reference simulation from Figure \ref{fig: collisionEvoSingleSetup}, and other lines show simulations that differ from the reference setup by the parameters listed in the key. At early times collisions make the edge slightly flatter (i.e. increase $\sigmaRI$), before making it significantly flatter as the time approaches $\tauMax(\rIZero)$. The local peak around $\tauMax$ should be ignored; it arises because the erf-powerlaw function used to fit the profile struggles with the complex edge shape at this time.}
    \label{fig: collisionEvoForTauMax}
\end{figure}

Figure \ref{fig: collisionEvoSingleSetup} was for one example setup. Figure \ref{fig: collisionEvoForTauMax} shows how the inner-edge steepness collisionally evolves in a number of different setups, demonstrating that this evolution is both qualitatively and quantitatively similar across a broad parameter space. Starting with the setup from Figure \ref{fig: collisionEvoSingleSetup} as a reference, we vary the input parameters and re-run the collisional simulation multiple times. We see that all of the setups have the same general collisional evolution. First, between $\tauB$ and $\tauMax$ the inner edge becomes slightly flatter, with $\sigmaRI$ increasing by a factor of ${\sim1.2}$. This remains constant, until the time approaches ${\tauMax}$; around this time the edge becomes much flatter, and tends towards the $r^2$ collisional profile. The plot shows the importance of $\tauMax$ in setting the collisional evolution of the sculpted-disc edge.

\subsubsection{Implications of collisions for planet-sculpted debris discs}
\label{subsec: collisionsImplications}

The previous sections showed that collisions have a very small effect on the inner-edge profile, until the time $\tauMax$. After this, a planet-sculpted edge would become significantly flatter. Consideration of $\tauMax$, the stellar lifetime and the planet-sculpting timescale $\tDiff$ are therefore vital for assessing the effect of collisions on the edge profile.

We can use the above timescales to identify three possibilities for the evolution of the debris-disc inner edge, depending on $\tDiff$ and $\tauMax$:

\begin{itemize}
    \item $\tDiff \ll \tauMax$: the planet sculpts the disc, producing a sharp inner edge like our $n$-body simulations. This edge then gradually gets flatter due to collisions, but does not change significantly until the time approaches $\tauMax$. Then, if left long enough, it may eventually assume an $r^2$ profile.
    \item $\tDiff \sim \tauMax$: unknown outcome. In this regime planetary sculpting and collisions are equally important at all times, and our simulations are insufficient to model this.
    \item $\tDiff \gg \tauMax$: the disc collisionally evolves almost as if no planet were present. The disc tends toward an $r^2$ profile.
\end{itemize}

\noindent So if $\tDiff$ and $\tauMax$ are known, then the relative importance of collisions to planetary sculpting can be ascertained, and the evolution of the disc edge predicted. However, whilst $\tDiff$ is straightforward, Equations \ref{eq: GInCollisionModel} to \ref{eq: tauMax} show that $\tauMax$ is a complicated function depending on many parameters, several of which are interdependent and unknown.

We can gain deeper insight by rewriting $\tauMax$, to make its dependencies more explicit. Equations \ref{eq: GInCollisionModel} to \ref{eq: tauMax} give $\tauMax$, $G$ and $\xC$ in their general forms; we can simplify these by substituting our assumed form for the initial $\SigmaR$, as well as $\qDStar$, $f$, $I/e$ etc. We also assume ${\sMax\gg\sMin}$. For the region beyond the inner edge, this yields

\begin{multline}
    \tau_{\rm max} = \frac{649 \yr}{\xM} \; \frac{\qG-5/3}{2-\qP} \left(\frac{r}{\au}\right)^{3/2+\alphaR} \left(\frac{\sMax}{\km}\right) \left(\frac{m_*}{\mSun}\right)^{-1/2}\\
    \times \left(\frac{\rho}{\gPerCmCubed}\right) G^{-1}(\qG, \sMax, r),
    \label{eq: tauMaxSimplified}
\end{multline}

\noindent where

\begin{multline}
    G(\qG, \sMax, r) = \left[\xC(\sMax, r)^{5-3\qG}-1\right]\\
    + 2\frac{\qG-5/3}{\qG-4/3}\left[\xC(\sMax, r)^{4-3\qG}-1 \right]\\
    + \frac{\qG-5/3}{\qG-1}\left[\xC(\sMax, r)^{3-3\qG}-1 \right]
    \label{eq: GInCollisionModelSimplified}
\end{multline}

\noindent and

\begin{equation}
    \xC(\sMax, r) = 0.0138 \; e^{-1/2} \left(\frac{\sMax}{\km}\right)^{\bG} \left(\frac{r}{\au}\right)^{1/4} \left(\frac{m_*}{\mSun}\right)^{-1/4}.
    \label{eq: XcInCollisionModelSimplified}
\end{equation}

\noindent Expressing the equations in this form demonstrates that $\tauMax$ is inversely proportional to the unknown $\xM$, which is related to the initial mass of the debris disc in MMSN units. We use this to better visualise $\tauMax$.

\begin{figure*}
	\includegraphics[width=17cm]{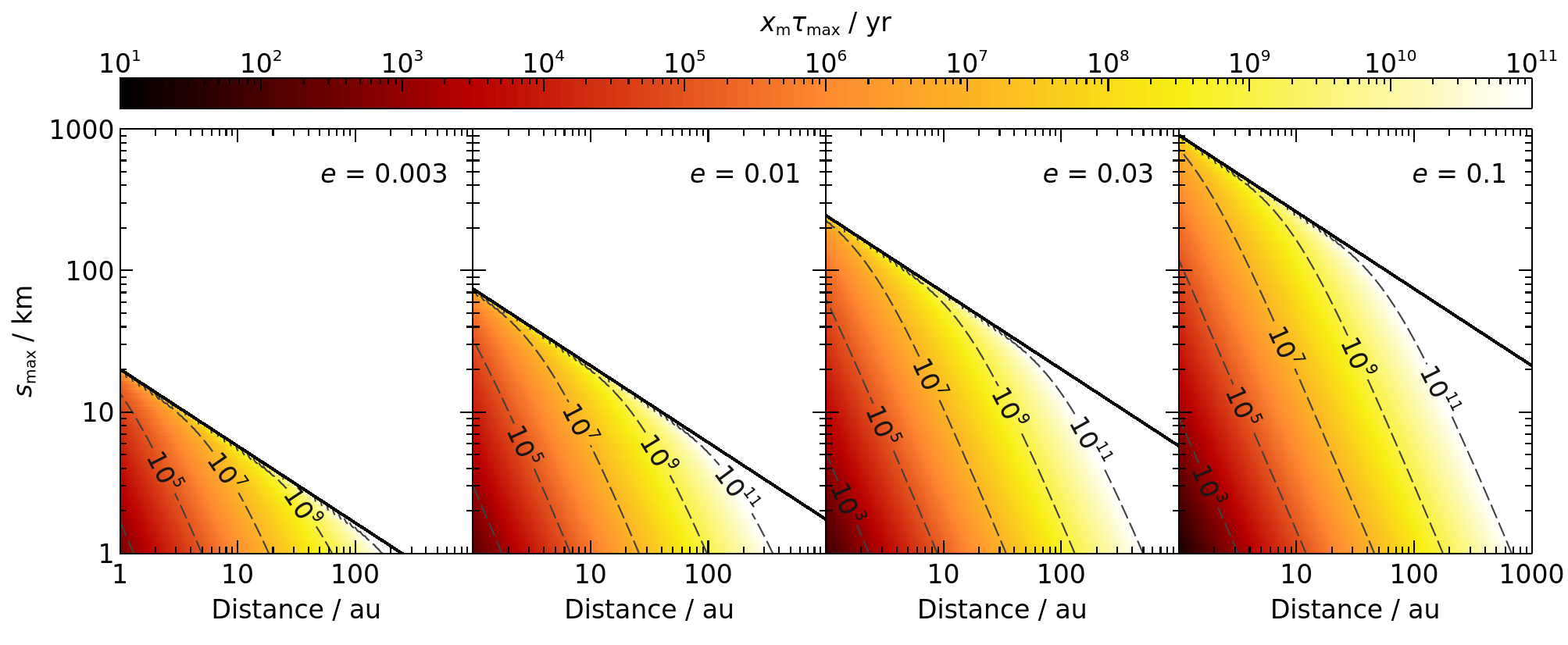}
    \caption{Plots of $\tauMax$ (collisional lifetime of the largest debris) multiplied by $\xM$ (initial surface density relative to the MMSN), for a Solar-type star and a disc with an initial surface-density profile of $r^{3/2}$. The plots show that $\tauMax$ strongly depends on distance, $\sMax$ and debris eccentricity, and can sometimes be very long. It is unlikely that ${\xM>10}$, so $\tauMax$ can be much longer than the ${\sim10^7}$ to ${10^{9}\yr}$ ages of debris-disc systems. If the largest bodies lie above the solid line (Equation \ref{eq: sMaxForTauMaxInfinite}), then they would never be destroyed and $\tauMax$ becomes infinite. Note that $\xM$ and $\sMax$ cannot be arbitrarily varied for an observed debris disc; only certain combinations are allowed, since they set the observed-dust mass at the system age (Equation \ref{eq: mDustAtT}). Increasing or decreasing the star mass respectively raises or lowers the contours slightly, but does not have a large effect.}
    \label{fig: tauMaxTimesXm}
\end{figure*}

Figure \ref{fig: tauMaxTimesXm} shows $\tauMax$ multiplied by $\xM$, for a Solar-type star and a disc with an initial surface-density profile of $r^{3/2}$. The figure shows that, for reasonable debris parameters, $\tauMax$ is typically extremely long, unless $\sMax$ is very small or $\xM$ very large. However, $\sMax$ is unlikely to be smaller than a few kilometers, otherwise it would violate planetesimal-formation models and the statistics of discs of various ages \citep[and refs therein]{KrivovWyatt2021}. A ${100\au}$ disc with debris eccentricity 0.01 would therefore have ${\xM \tauMax \sim 10^{10} \yr}$. Furthermore, ${\xM>10}$ is unlikely because such debris discs would have masses exceeding those of solids in the preceding, protoplanetary stage \citep{KrivovWyatt2021}. As a result, for many observed debris discs $\tauMax$ may be longer than the ${\sim10^7}$ to ${10^{9}\yr}$ system age. In fact, ${\tauMax}$ can even be infinite, which means that collisions never disrupt the largest bodies; this occurs in the large regions of parameter space where ${\xC(\sMax) \geq 1}$ (for which ${G\leq0}$). Rearranging Equation \ref{eq: XcInCollisionModel}, $\tauMax$ becomes infinite if

\begin{equation}
    \qDStar(\sMax) \geq \frac{f^2 G m_*}{2r},
    \label{eq: sMaxForTauMaxInfinite}
\end{equation}

\noindent which yields the solid lines on Figure \ref{fig: tauMaxTimesXm}.

Conversely, calculating $\tDiff$ (Equation \ref{eq: diffusionTime}) shows that a Jupiter-mass planet located within ${100\au}$ of a Solar-mass star would finish sculpting a disc within ${<10\myr}$. So if such planets were sculpting debris discs, then they would likely do so long before collisions had any effect on the inner edges. This would make the inner-edge profiles long lived, and since even ${10\tauMax}$ is insufficient to fully erase the planet's signature in the example on Figure \ref{fig: collisionEvoSingleSetup}, it is likely that a disc sculpted by a massive planet would maintain a distinctive shape even if collisions were fully accounted for. For such cases, the planet can be constrained using the process in Section \ref{subsec: modelExampleFit}. For smaller, Neptune-mass planets the sculpting timescale is 100 times longer, in which case both collisions and planets could simultaneously sculpt the discs. Such systems may not have reached their final configurations by the time they are observed, meaning that the disc edges would be somewhere between the very steep values expected for such planets, and the $r^2$ profile expected from collisions. This is discussed further in \mbox{Section \ref{subsec: discussionWhyObsEdgesFlatterThanSims}}.

\section{Application to observed debris discs}
\label{sec: applicationToObservedDiscs}

Finally, we now compare the inner-edge slopes we predict from planetary sculpting to the inner edges of observed debris discs. We consider seven ALMA-resolved, extrasolar debris discs, each of which has a literature measurement of its inner-edge steepness. The systems are listed in Table \ref{tab: applicationToRealSystems}, and were analysed by \cite{Lovell2021}, \cite{Faramaz2021} and \cite{ImazBlanco2023}.

\begin{table*}
	\centering
	\caption{Constraints on the inner edges of several axisymmetric, extrasolar debris discs. The parameters $\rT$, $\pI$, $\pO$ and $\eta$ are from literature fits to the surface-density profiles in the inner-edge regions (Equation \ref{eq: lovellSDProfile}), and $\rI$ and $\sigmaRI$ are the equivalent parameters in our erf-powerlaw model for comparison with our simulations (Equation \ref{eq: fittedProfileForLitComparison}). Note that $\pI$ and $\pO$ are often 0.5 larger than the values provided in disc-imaging papers, because we consider the surface-density slope whilst those papers often give the intensity slope (e.g. \citealt{ImazBlanco2023}).}
	\label{tab: applicationToRealSystems}
	\begin{tabular}{r c r r r r r r r r c} 
		\hline
            \multicolumn{1}{c}{System} & \multicolumn{1}{c}{Name} & \multicolumn{1}{c}{Star mass /$\mSun$} & \multicolumn{1}{c}{Age /${\myr}$} & \multicolumn{1}{c}{$\rT \; /\au$} & \multicolumn{1}{c}{$\pI$} & \multicolumn{1}{c}{$\pO$} & \multicolumn{1}{c}{$\eta$} & \multicolumn{1}{c}{$\rI \; /\au$} & \multicolumn{1}{c}{$\sigmaRI$} & \multicolumn{1}{c}{Refs.} \\
		\hline
            HD 9672 & 49 Ceti & $1.98\pm0.01$ & $45\pm5$ & $131^{+13}_{-12}$ & $1.3^{+0.4}_{-0.3}$ & $-3.0^{+0.4}_{-0.5}$ & 2 & \multicolumn{1}{c}{-} & $0.6^{+0.1}_{-0.2}$ & 1, 3 \\
            HD 10647 & q$^1$ Eri & $1.13^{+0.03}_{-0.04}$ & $1700\pm600$ & $76\pm1$ & $>5.6$ & $-2.1\pm0.2$ & 2 & 70 & $<0.14$ & 1, 4\\
            HD 92945 & V419 Hya & $0.87\pm0.01$ & $250\pm100$ & $54\pm2$ & $8\pm2$ & $-0.8^{+0.4}_{-0.6}$ & 2 & 52 & $0.11\pm0.03$ & 1, 3 \\
            HD 107146 & \multicolumn{1}{c}{-} & $1.03^{+0.02}_{-0.04}$ & $150^{+100}_{-50}$ & $44\pm2$ & $7.2^{+0.9}_{-0.7}$ & $-0.2^{+0.1}_{-0.2}$ & $2.8^{+1.2}_{-0.7}$ & 41 & ${0.11\pm0.01}$ & 1, 3 \\
            HD 197481 & AU Mic & $0.59\pm0.03$ & $24\pm3$ & $36.4\pm0.7$ & $1.4\pm0.4$ & $-9\pm1$ & 2 & \multicolumn{1}{c}{-} & $0.6\pm0.3$ & 1, 3 \\
            HD 206893 & \multicolumn{1}{c}{-} & $1.32^{+0.07}_{-0.05}$ & $160\pm20$ & $35^{+7}_{-10}$ & $>1.1$ & $-2.2\pm0.2$ & 2 & \multicolumn{1}{c}{-} & $<0.76$ & 2, 3 \\
            HD 218396 & HR 8799 & $1.59\pm0.02$ & $42^{+6}_{-4}$ & $180^{+10}_{-20}$ & $3.0^{+0.9}_{-0.5}$ & $-0.6^{+0.5}_{-0.3}$ & 2 & 160 & $0.27^{+0.04}_{-0.08}$ & 1, 5\\
		\hline
            \multicolumn{11}{l}{References: Star masses and ages from (1) \citet{Pearce2022ISPY}, (2) \citet{Hinkley2023}. ALMA surface-density fits from  (3) \citet{ImazBlanco2023},}\\
            \multicolumn{11}{l}{(4) \citet{Lovell2021}, (5) Model 2 in \citet{Faramaz2021}.}
	\end{tabular}
\end{table*}

The literature works used various parameterisations to quantify disc profiles, which we must first convert to erf-powerlaw functions for comparison with our simulations. The literature works generally fitted the inner edges using a radial powerlaw, with some turnover occurring further out; this turnover could be the disc outer edge, or a gap within the disc. For our analyses the nature of the turnover is unimportant, but its location is needed to define the `inner' region of the disc. Following \citet{Lovell2021}, we quantify the surface densities around the inner edges of the literature discs using a double powerlaw:

\begin{equation}
    \SigmaR \propto \left[\left(\frac{r}{\rT}\right)^{-\eta \pI} + \left(\frac{r}{\rT}\right)^{-\eta\pO}\right]^{-1/\eta},
    \label{eq: lovellSDProfile}
\end{equation}

\noindent where $\pI$ is the slope of the inner edge, $\rT$ is the radial location of the turnover, $\pO$ is the slope exterior to the turnover, and $\eta$ sets how sharp the turnover is. The assumed values for each disc are listed in Table \ref{tab: applicationToRealSystems}; for fits where $\eta$ was undefined or unconstrained, we assumed a value of 2 as in \citet{Lovell2021}.

We next convert these powerlaw fits into erf-powerlaw profiles for comparison with our simulations. This is not strictly valid as we should really fit our erf-powerlaw profile to the underlying data, but it should be sufficient for this simple analysis. We consider the profile

\begin{equation}
    \SigmaR \propto \left[ 1-{\rm erf}\left(\frac{\rI - r}{\sqrt{2} \sigmaRI \rI} \right) \right] \left(\frac{r}{\rI} \right)^{\pO},
    \label{eq: fittedProfileForLitComparison}
\end{equation}

\noindent which is equivalent our \mbox{Equation \ref{eq: fittedProfile}} near the inner edge. We set $\pO$ to the literature values describing the region just beyond the edge, noting that $\pO$ is typically negative. We then use Equation \ref{eq: sigmaRIToBeSteeperThanPowerlaw} to convert the inner-edge powerlaw index $\pI$ to an equivalent flatness $\sigmaRI$, which approximates how the inner edge would appear if fitted with an erf function instead. Finally, we deduce $\rI$, which characterises the location of the inner edge in the erf model, such that Equations \ref{eq: lovellSDProfile} and \ref{eq: fittedProfileForLitComparison} peak at the same radial location. A comparison of the erf-powerlaw and double-powerlaw profiles for an example system is shown on Figure \ref{fig: q1EriModelComparison}, and the inferred values of $\sigmaRI$ and $\rI$ are listed for all systems in Table \ref{tab: applicationToRealSystems}. In some cases $\rI$ could not be fitted because the region beyond the turnover was so steep that Equation \ref{eq: fittedProfileForLitComparison} does not turn over; however, this does not affect the edge flatness $\sigmaRI$, which is the parameter of interest in this section.

\begin{figure}
	\includegraphics[width=8cm]{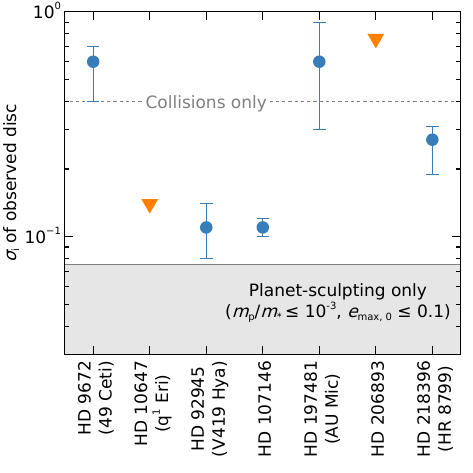}
    \caption{Flatness of the inner edges of observed discs (Table \ref{tab: applicationToRealSystems}). Blue circles and orange triangles are $\sigmaRI$ values and upper limits respectively. Several discs (\mbox{49 Ceti}, \mbox{AU Mic} and \mbox{HD 206893}) have flat inner-edge slopes consistent with pure collisional evolution, as shown by the dashed horizontal line. Other discs (\mbox{q$^1$ Eri}, \mbox{HD 92945} and \mbox{HD 107146}) have much steeper profiles, but these are still flatter than expected from \textit{in-situ} sculpting of low-eccentricity discs by circular-orbit planets.}
    \label{fig: obsSteepnessVsTheory}
\end{figure}

Figure \ref{fig: obsSteepnessVsTheory} shows the inferred $\sigmaRI$ values for the observed discs, which exhibit a range of inner-edge slopes. At least two have relatively flat slopes consistent with pure collisional evolution, which would yield powerlaw indices of ${\pI\approx2}$ as described in Section \ref{sec: collisions} (equivalent to ${\sigmaRI\approx0.4}$, shown as the dashed line on Figure \ref{fig: obsSteepnessVsTheory}). Several have much steeper inner edges, which \cite{ImazBlanco2023} argued are indicative of planetary sculpting. However, whilst those edges are indeed steeper than expected from collisions alone, they are still \textit{flatter} than our simulations predict for \textit{in-situ} sculpting of low-eccentricity discs by planets on circular orbits. The shaded region on Figure \ref{fig: obsSteepnessVsTheory} shows the maximum $\sigmaRI$ expected if a ${\mPOverMStar \leq 10^{-3}}$ planet sculpts a disc with ${\eMaxZero \leq 0.1}$; all of the observed inner edges are flatter than this\footnote{Whilst a planet with mass ${\mPOverMStar \approx 0.07}$ could drive $\sigmaRI$ to required values of ${0.11}$ (Figure \ref{fig: sigmaRIn}), we disfavour this possibility because it seems unlikely that planets with such specific masses exist in two of our seven systems.}. In Section \ref{subsec: discussionWhyObsEdgesFlatterThanSims} we discuss the potential implications of these flatter edges, which could teach us about architectures, dynamical processes and histories in the outer regions of planetary systems.

\section{Discussion}
\label{sec: discussion}

In Section \ref{subsec: discussionWhyObsEdgesFlatterThanSims} we discuss the possible implications of observed discs having flatter-than-expected edge profiles, and in Section \ref{subsec: discussionComparisonToPreviousStudies} we compare the analyses in this paper to previous studies.

\subsection{Why are observed inner edges flatter than expected from sculpting planets?}
\label{subsec: discussionWhyObsEdgesFlatterThanSims}

In Section \ref{sec: applicationToObservedDiscs} we showed that, whilst the edges of several ALMA-resolved discs are steeper than expected from collisions alone, they are flatter than expected from planetary sculpting of an initially low-eccentricity disc. Here we discuss several possible reasons for this, and what these could teach us about planetary systems.

\subsubsection{Debris discs have high intrinsic-excitation levels}

One possibility is that planetesimals in debris discs have higher intrinsic excitation levels than is often assumed. For example, if debris at a disc inner edge had ${\eRMSI \approx 0.13}$, then that edge would have ${\sigmaRI\approx0.11}$, which is similar to the steepest discs in Table \ref{tab: applicationToRealSystems}. This value of $\eRMSI$ is higher than the maximum of 0.06 that could be imparted by a non-migrating, circular-orbit planet (Figure \ref{fig: eRMSI}), raising the possibility that these eccentricities arise through non-planetary processes. Examples of possible excitation mechanisms include self stirring by the debris disc's self gravity, or processes that occurred during system formation. \cite{Marino2021} used the \textit{outer} edges of several observed discs to infer the debris-excitation levels, and found these to be consistent with ${\eRMSI \gtrsim 0.1}$; that sample included ${\HD92945}$ and ${\HD107146}$, which have the steepest edges in Table \ref{tab: applicationToRealSystems}. It is therefore plausible that \textit{in-situ} planets sculpted the inner edges of some observed discs, but that this debris has separately been excited by some non-planetary processes.

One means to test this would be to measure scale heights at the \textit{inner} edges of debris discs. We find that MMRs of a coplanar planet excite debris eccentricity but not inclination (Section \ref{subsec: simulationsExamples}), so if \textit{in situ}, coplanar planets sculpt debris-disc inner edges, then these edges would be more excited radially than vertically.

\subsubsection{Planets are present at the inner edges, but sculpting has not yet finished}

Planets take time to eject debris and fully sculpt a debris-disc inner edge. The required time is characterised by 1 to 10 diffusion timescales (Section \ref{subsec: simulationsSDTimeEvolution}). Until this time, the disc's inner edge could have almost any profile, which would depend on its initial configuration. It is therefore possible that planets reside just interior to debris discs, but the planets have not yet finished sculpting the edges into steep profiles. This is particularly plausible for young systems, or those with low-mass planets.

The observed discs in Table \ref{tab: applicationToRealSystems} with the flattest inner-edge profiles are also the youngest, with ages less than ${100\myr}$. These are \mbox{49 Ceti}, \mbox{AU Mic} and ${\HR 8799}$. Equation \ref{eq: minMPMassFromDiffusionTime} shows that, if planets with masses below 0.1 to ${1\mJup}$ lie at the inner edges of those discs, then these systems would be younger than one diffusion time and hence planetary sculpting would not yet have finished. It is therefore plausible that inner-edge profiles are flatter than expected not because sculpting planets are absent, but because sculpting has not yet finished. If true, then care must be taken when inferring planet properties from debris discs, because the hypothesised planet-disc interactions may not have reached their final state. This may be a prevalent problem because the discs most favourable for observation are generally the youngest, for which the dust mass (and hence brightness) is highest.

\subsubsection{Discs are sculpted by migrating planets}

In this paper we modelled planets on non-evolving orbits, but a migrating planet would impose a very different morphology on the disc inner edge. Specifically, a migrating planet causes MMR sweeping, where the nominal locations of MMRs move across the debris disc and trap large numbers of bodies in resonance. During this process, resonant debris is excited to increasingly high eccentricities (e.g. \citealt{Wyatt2003, Reche2008, Friebe2022}), which would result in a flatter edge profile.

An example of this occurred in the Solar System. Neptune is located at the inner edge of the Kuiper Belt, but the orbits of Kuiper-Belt objects (KBOs) show that the planet migrated outwards in the past \citep{Malhotra1993}. As a result, KBOs at the Kuiper Belt's inner edge have relatively high eccentricities, which make the Belt's inner edge flatter than would be expected from perturbations by a non-migrating Neptune. In an upcoming paper (Morgner et al., in prep.), we take the observed KBOs and, using a method similar to \cite{Vitense2010}, de-bias these observations to estimate the true KBO population distribution. We then fit the surface density of this de-biased population with Equation \ref{eq: fittedProfileForLitComparison}, to get the steepness of the Kuiper-Belt inner edge. This yields ${\sigmaRI=0.076}$, which is much flatter than would be expected for a non-migrating Neptune; Neptune has ${\mPOverMStar=5\times10^{-5}}$, so the inner edge should have ${\sigmaRI < 0.02}$ if Neptune's orbit never changed (Figure \ref{fig: sigmaRIn}). The reason for this difference is that Neptune excited KBOs through MMR sweeping as it migrated, resulting in eccentricities at the Kuiper Belt's inner edge of ${\eRMSI\sim0.2}$, compared to just 0.01 expected from excitation by a non-migrating Neptune (Equation \ref{eq: planetERMSFit}).

It is possible that some extrasolar discs were also sculpted by migrating planets, which made their edge profiles flatter than our \textit{in-situ} model predicts. Dust clumps that may be indicative of planetary migration are inferred in debris discs \citep{Lovell2021, Booth2023}, and \cite{Pearce2022ISPY} argue that migration may be required to relate the location of debris-disc edges to system-formation theory. These results, combined with debris-disc edges being flatter than expected for sculpting by non-migrating planets, could imply that planetary migration is common in the outer regions of debris-disc systems. Such migration could be caused by planetesimal scattering (e.g. \citealt{Friebe2022}), planet-gas interactions in the protoplanetary disc phase, or planet-planet scattering. 

It may be possible to test exoplanet-migration scenarios in the near future. \textit{JWST} will search for planets in the outer regions of debris-disc systems, and should detect many of the inferred planets if they exist \citep{Pearce2022ISPY}. If a planet were found just interior to a debris disc, and that disc had a flatter edge profile than would be expected from \textit{in-situ} planetary sculpting, then it could be evidence that the planet migrated outwards in the past, exciting debris as it did so.

\subsubsection{Collisions have flattened the planet-sculpted edges}

Debris-disc edges could initially have been sculpted by planets, and these edges could since have undergone significant collisional evolution. The steepest edges of the Table \ref{tab: applicationToRealSystems} discs have ${\sigmaRI\approx0.11}$, which is at least 1.3 times flatter than expected from sculpting by non-migrating planets (Figure \ref{fig: sigmaRIn}). However, Figure \ref{fig: collisionEvoForTauMax} shows that collisions can quickly increase $\sigmaRI$ by a factor of 1.2, long before the largest bodies start colliding at $\tauMax$. This could explain the steepest inner edges in Table \ref{tab: applicationToRealSystems}; in these cases, the system ages could be longer than $\tauB$ but shorter than $\tauMax$. After $\tauMax$, collisions would flatten the edge further, increasing $\sigmaRI$ by a factor of 3 to 6 within $10\tauMax$; this could potentially reproduce some of the flatter edges in Table \ref{tab: applicationToRealSystems}, if those systems are already older than $\tauMax$. Since some observed edges are too steep to be explained by collisions alone, yet too flat to be explained by non-migrating planets alone, this could suggest that both processes play a significant role in some systems. 

\subsubsection{Planets are present at debris-disc inner edges, but the edges are set by planet formation rather than scattering}

A circular-orbit planet would eject debris that initially lies within \mbox{3 Hill} radii of its orbit (Figure \ref{fig: rIn_ap}). Therefore, if a newly formed system had a planetesimal disc extending down to less than \mbox{3 Hill} radii exterior to a planet's orbit, then that planet would eventually impart a sharp inner edge on the disc. This is the scenario we modelled in our simulations. However, a real planet would also have accreted material from around its orbit as it formed. If the forming planet's feeding zone were wider than \mbox{3 Hill} radii, then much of the debris within the scattering radius would already have been accreted by the time the planet formed. Also, other processes in the protoplanetary disc phase could further affect planetesimal profiles, such as the forming and trapping of material in planet-induced pressure bumps near a planet. Therefore, it is possible that a debris disc's edge profile is set by processes occurring during planet formation, rather than by pure planetary scattering; in this case, its edge profile could differ from those that we model (e.g. \citealt{Eriksson2020}).

\subsubsection{Debris-disc edges are set by planetesimal formation alone}
\label{subsec: discussionEdgesFromPltmlFormation}

Planets are often invoked to explain the shapes and locations of debris-disc inner edges, motivated in part by Neptune dominating the inner edge of the Kuiper Belt. However, an alternative possibility is that planetesimals naturally form in distinct radial zones; in this case debris-disc edges may be completely unrelated to planets, but instead mark locations were planetesimal formation transitioned from being inefficient to efficient. Such formation zones are predicted by some streaming-instability models \citep{Carrera2017}, as well as models involving snowlines \citep{Ida2016, Schoonenberg2017, Drazkowska2017, Schoonenberg2018, Izidoro2022, Morbidelli2022}. This possibility can soon be tested because, if sculpting planets exist, then many should be detectable by \textit{JWST} \citep{Pearce2022ISPY}; the absence of such detections would potentially imply that other processes set disc-edge locations.

\subsubsection{Debris discs do not have sharp cutoffs in semimajor axis}

Debris-disc inner edges could be broad in semimajor-axes space, meaning that the number density of debris gradually increases with semimajor axis. The radial profile of such a disc would mimic that of a higher-eccentricity disc with a sharper semimajor-axis distribution. Shallow semimajor-axis distributions could be a natural outcome of system formation (Section \ref{subsec: discussionEdgesFromPltmlFormation}), arise through MMR sweeping during planet migration \citep{Friebe2022}, or result from debris diffusion during self stirring \citep{Ida1993}. Since the average eccentricities and inclinations would be related in a relaxed debris disc, the disc vertical thickness would test this; \cite{Marino2021} argued that the outer region of the AU Mic disc is more extended radially than vertically, which could imply low eccentricities and hence a slowly changing semimajor-axis profile.

\subsubsection{Debris discs are sculpted by planets with low but non-zero eccentricities}

Eccentric planets can excite debris to higher eccentricities than circular-orbit planets, due to additional secular perturbations (eccentric planets are considered in Appendix \ref{app: eccentricPlanets}). This would result in flatter disc inner edges. Combining Equations \ref{eq: rSigInAnalyticEstimate} and \ref{eq: eMaxSec}, we can show that a Neptune-mass planet orbiting a solar-type star with eccentricity 0.10 could induce an inner-edge steepness of ${\sigmaRI \approx 0.11}$, like the sharpest observed edges. Such a planet would make the inner edge elliptical with global eccentricity 0.052 (Equation \ref{eq: eInnerEdge}), which is high enough to detect with modern observations (e.g. \citealt{Lovell2021}). Therefore, it should be possible to rule out an eccentric planet as being responsible for a flat disc edge if the disc's global eccentricity were robustly measured as negligible.

However, an intriguing possibility arises if the star position is poorly constrained. In this scenario, a disc with a low global eccentricity could masquerade as axisymmetric, owing to a poorly constrained stellar offset. Were this the case, then an eccentric planet could flatten the disc inner edge, whilst no global disc eccentricity would be detected. One way to rule out this scenario would be to measure the azimuthal variation in the disc profile, because a planet sculpting a truly eccentric disc would make the inner edge flatter at pericentre than apocentre (Section \ref{subsec: eccentricPlanetQuantify}). Depending on the disc width and the intrinsic debris excitation, the eccentric planet may also flatten the outer edge, which could be observable (compare Figures \ref{fig: lowESurfaceDensities} and \ref{fig: appEccentricPlanet}).

\subsubsection{Unresolved Trojans are present interior to the inner edges}
\label{subsec: discussionTrojans}

Unresolved Trojans could be present at the disc inner edges. In our simulations, we purposefully truncated the initial-disc edge at one Hill radius exterior to the planet, to omit any Trojans and ensure that our fitted ${\sigmaRI}$ values describe the `true' inner-edge slopes. However, if sculpting planets are present at the inner edges of real discs, but there are also unresolved Trojans co-orbiting with the planets, then the Trojans would flatten the azimuthally averaged inner-edge profiles. The degree of flattening would depend on the surface density of Trojans relative to the disc.

\subsubsection{Dust transport is important}

Dust-transport processes could be operating in addition to planetary sculpting, which would make the disc edges more radially extended and their profiles flatter. Poynting-Robertson (PR) drag can be significant for small grains, causing discs to be radially extended in scattered light; whilst the larger, ALMA-imaged grains we consider should be unaffected by \mbox{PR drag}, they could be affected in a similar way by stellar winds. Winds would be particularly important for discs around late-type stars, and would drag material inwards from the planetesimal belt (e.g. \citealt{Plavchan2005, Reidemeister2011, Schuppler2015}). This would flatten the inner-edge profile. Similarly, CO gas is detected in many debris discs around early-type stars, which could also cause radial drift through gas drag (e.g. \citealt{Krivov2009, Marino2020, Pearce2020}). Again, this would flatten the inner edges.

\subsubsection{More complex disc-planet interactions occur in some debris discs}

The inner edges of some specific systems in Table \ref{tab: applicationToRealSystems} could be excited by more complex planet-disc interactions. There are gaps in the discs of both ${\HD92945}$ and ${\HD107146}$, which are the discs with the steepest inner edges. \cite{Friebe2022} showed that the simplest and most self-consistent way to explain the morphology of ${\HD107146}$ is if a planet has migrated across the gap; such migration would cause sweeping by mean-motion resonances, which would excite debris at the disc inner edge (their Figure 9, lower-right panel). In that model the planet would now lie just exterior to the inner edge of the gap, and may have swept up a Trojan population which could resemble the additional gap features in \cite{ImazBlanco2023}. It is therefore possible that planets lying further out in more-complex discs could excite debris at the inner edge, leading to flatter edge profiles.

Alternatively, planets could excite inner-edge debris through secular resonances rather than mean-motion resonances, which could lead to higher debris eccentricities and thus flatter edges. A secular resonance occurs where the apsidal precession rate of debris (due to the planet and disc self-gravity) matches that of the planet (due to the disc or other planets), and such resonances can drive up debris eccentricities and even open remote gaps in broad discs \citep{Pearce2015HD107146, Yelverton2018, Sefilian2021, Sefilian2023}. 
If a secular resonance were located near the disc inner edge, then it could dominate over scattering and MMRs, and hence produce a different edge location and profile to those in our simulations (e.g. \citealt{Smallwood2023}). Given the very large uncertainties on both debris-disc masses and orbital architectures in the outer regions of systems, we cannot state for certain whether this secular-resonance excitation occurs, but we can estimate the disc masses that could lead to this scenario.

In a coplanar system with a single planet and an external, self-gravitating disc, a secular resonance occurs at semimajor axis $\aSec$. The location of $\aSec$ depends on the total disc mass $M_{\rm d}$ relative to that of the planet, as well as the planetary semimajor axis. A general expression describing this relationship is derived in \citet{Sefilian2021} (their Equation 19) which, assuming a power-law disc with surface density $\propto a^{-1.5}$ spanning semimajor axes $a_{\rm in}$ to $a_{\rm out}$, with ${\aP \approx a_{\rm in}}$, reads as

\begin{equation}
 \frac{M_{\rm d}}{\mP} \approx \frac{1.5}{|\psi_1(\aSec)|} \left(\frac{a_{\rm out}}{a_{\rm in}}\right)^{1/2} \left( \frac{\aSec}{a_{\rm in}}  \right)^{-2.5} , 
 \label{eq: res_condition_AS}
\end{equation}

\noindent where $\psi_1(a)$ is a factor of order unity (see also \citealt{Sefilian2023}). A secular resonance will hence lie near the disc inner edge if ${M_{\rm d}/\mP \gtrsim 0.89 \left(a_{\rm out}/a_{\rm in}\right)^{1/2}}$; here we arbitrarily define `near' as ${\aSec < 1.2 a_{\rm in}}$, where ${\psi_1(1.2 a_{\rm in}) \approx -1.07}$ (computed using Equation 18 in \citealt{Sefilian2019}, assuming  ${H = 0.05}$ for the disc scale height\footnote{For comparison, defining the inner-edge region as ${< 1.5 a_{\rm in}}$ instead yields ${M_{\rm d}/\mP \gtrsim 0.75 \left(a_{\rm out}/a_{\rm in}\right)^{1/2}}$, with ${\psi_1(1.5 a_{\rm in}) \approx -0.73}$.}). This implies that, for disc-to-planet mass ratios of roughly unity or higher, secular resonances could play a role in shaping the disc inner edge. Since debris discs could have masses up to ${1000\mEarth}$ \citep{KrivovWyatt2021}, this effect could be significant even for Jupiter-mass planets. We will further investigate the effect of secular resonances on debris-disc edges in a future work (Sefilian et al., in prep.). 

\subsection{Comparison to literature studies}
\label{subsec: discussionComparisonToPreviousStudies}

Several literature studies also assessed the effect of a sculpting planet on debris-disc inner edges. These used various different approaches, and yielded various different results. In this Section we compare our paper to literature works, specifically focussing on the $n$-body results (Section \ref{subsec: discussionComparisonToPreviousStudiesGravity}), collisional modelling (Section \ref{subsec: discussionComparisonToPreviousStudiesCollisions}) and the planet-inferring technique of \cite{Pearce2022ISPY}, which uses the locations of debris-disc inner edges but not their steepnesses (Section \ref{subsec: discussionComparisonToPreviousStudiesPlanetPars}).

\subsubsection{Gravitational effects}
\label{subsec: discussionComparisonToPreviousStudiesGravity}

The fact that MMRs of planets on circular orbits excite debris at disc inner edges, and that more-massive planets cause higher excitation and hence flatter edge profiles, was also identified in previous works. \cite{Mustill2012} used encounter maps to demonstrate this effect on planetesimals (their Figure 4), and showed that a similar effect arises in the prescription of \cite{Wisdom1980}. \cite{Chiang2009} and \cite{Rodigas2014} consider small dust released from a population of planet-sculpted planetesimals, and show that this decrease in edge steepness with increasing planet mass is also expected in scattered-light observations (their Figures 3 and 1b respectively). 

\cite{Quillen2006Fom} and \cite{QuillenFaber2006} used MMR theory to predict that a low-eccentricity planet should impose an eccentricity dispersion at the disc inner edge that is proportional to $(\mPOverMStar)^{3/7}$, which is slightly steeper than our empirical $(\mPOverMStar)^{1/3}$ for a circular-orbit planet (Equation \ref{eq: planetERMSFit}). However, we note that the zero-planet-eccentricity simulations of \cite{QuillenFaber2006} appear more consistent with our flatter index of 1/3 (filled pentagons on their Figure 2), and this index is also predicted theoretically by \cite{Petrovich2013} (their Equations 19 and 34). We therefore argue that $(\mPOverMStar)^{1/3}$ is a better estimate of debris eccentricities at the inner edge of a disc sculpted by a circular, non-migrating planet, provided that ${\mPOverMStar \lesssim 3.3\times10^{-3}}$. Above this mass ratio, we observe a turnover where increasing the planet mass no longer excites debris to higher eccentricities; this is due to the inner edge of the planet's chaotic zone extending beyond the location of strong MMRs, and was not observed by the above authors because their simulations did not extend to such high mass ratios.

\cite{Tabeshian2016,Tabeshian2017} modelled the interaction between a planet and a planetesimal disc, and showed that the MMRs of a circular-orbit planet can impose asymmetric structure on an external disc (Figure 5 in \citealt{Tabeshian2016}). In particular, they show that MMRs can carve crescent-shaped gaps in the disc, and that the inner-edge profile can vary with azimuth. We find the same result; in the simulation on Figure \ref{fig: lowESimPos}, a crescent-shaped gap can be seen aligned with the planet, and the inner-edge steepness differs slightly on the left and right sides of the disc. However, we do not investigate this asymmetry in more detail, because the difference is small enough that it is unlikely to significantly affect any inferred-planet parameters. Also, \cite{Tabeshian2016} show that the degree of asymmetry scales with the planet-to-star mass ratio, and this asymmetry is already small for the comparatively high mass ratio on Figure \ref{fig: lowESimPos}; for smaller mass ratios, the asymmetry would be even less pronounced.

\subsubsection{Collisional effects}
\label{subsec: discussionComparisonToPreviousStudiesCollisions}

We find that collisions make the disc edge flatter, and move its characteristic radius outwards (Figures \ref{fig: collisionEvoSingleSetup} and \ref{fig: collisionEvoForTauMax}). This is in qualitative agreement with Figure 3 of \cite{Nesvold2015}, which shows that the same effects manifest in the {\sc smack} collision prescription; that prescription simultaneously models the collisional and \textit{n}-body dynamical evolution of debris \citep{Nesvold2013}. \cite{Nesvold2015} also find that it takes 10 to 100 collisional timescales for this process to occur (their Figure 4). However, our collisional models, and our definitions of collisional timescales, are fundamentally different. 

We employ the analytical model of \cite{Lohne2008}, where the most relevant collisional timescale is $\tauMax$, the collisional lifetime of the largest bodies. This predominantly depends on disc location, debris eccentricity, the disc's initial mass and the size of the largest bodies. The latter two are unknown, but for physically plausible values, a disc with inner edge at ${50\au}$ would have $\tauMax$ of at least ${\sim10^{7}\yr}$, and potentially much longer (Figure \ref{fig: tauMaxTimesXm}). This means that $\tauMax$ should be at least comparable to the ages of debris-disc stars (Table \ref{tab: applicationToRealSystems}), and may be much longer; based on this, we argue that collisions should \textit{not} significantly reduce the steepness of planet-sculpted edges in many observed discs.

This conclusion differs from \cite{Nesvold2015}, who suggest that collisions typically \textit{would} have a significant effect on the observed edges of planet-sculpted discs. They model the collisional cascade in detail and self-consistently with the dynamical evolution, unlike our simulations. However, the difference between our conclusions is unlikely to be due to our separation of dynamics and collisions, but rather differences between our collisional assumptions and those of \cite{Nesvold2015}.

There are two main reasons for this difference. First, the maximum grain radius in \cite{Nesvold2015} is ${\sMax = 10\cm}$ (their Table 1); these pebbles are depleted much faster than the ${\gtrsim1\km}$ planetesimals in our model, so their collisional erosion occurs on much shorter timescales. Second, the amount of dust in the \cite{Nesvold2015} models is higher because they assume a greater minimum grain size; extrapolating from their ${\sMin = 1\mm}$ to the typical ${\sMin = 1}$ to ${10\um}$ expected from radiation pressure blowout corresponds to vertical optical depths that are higher than ours by factors of 10 to 30, assuming a size distribution index of 3.5. 

Accounting for the respective differences in $\sMin$ and $\sMax$ each increase the ratio between the collision timescales of smallest dust (their Eq. 3) and that of the biggest objects (which govern the long-term evolution). \cite{Nesvold2015} define smallest-grain collisional timescales of $10^3$ to ${10^6\yr}$ for a sample of observed discs (their Section 4); instead of the factors of 10 to 100 in their Figure 4, planetesimals would evolve collisionally on timescales that are longer by several orders of magnitude, in line with our results. So this difference in timescales appears to be the cause of our different conclusions regarding the relative effects of planets and collisions on the disc inner edge.

Nonetheless, the \cite{Nesvold2015} model captures additional physics that is omitted in our simpler prescription. In particular, our model would not capture the expected collisional enhancement in resonant populations \citep{Stark2009}; this means we probably underestimate collision rates at the MMR-dominated inner edges, and therefore overestimate collisional timescales. This effect could be significant in some cases, because the inner-edge eccentricities in our low-eccentricity planet simulations are determined by MMRs. Detailed, self-consistent modelling is therefore required in future, to properly assess the balance between collisions and planet-debris interactions at the inner edges of debris discs.

\subsubsection{Comparison to \citet{Pearce2022ISPY} planet predictions}
\label{subsec: discussionComparisonToPreviousStudiesPlanetPars}

\cite{Pearce2022ISPY} give a general model to infer the minimum mass, maximum semimajor axis and minimum eccentricity of a perturbing planet based on the shape and location of a debris disc's inner edge. Those predictions are based on \cite{Pearce2014}, and do not consider the steepness of the disc edge; as a result they have a degeneracy in planet mass and semimajor axis, because they cannot distinguish a low-mass planet near the disc from a high-mass planet away from the disc. However, in this paper we make predictions that also include information about the edge steepness; this can break the degeneracy, because edge steepness can be directly related to planet mass (at least in the collisionless regime).

Section \ref{sec: simpleAnalyticModel} describes our simple analytic model to predict planet properties from an observed-disc edge. These predictions are comparable to those of \cite{Pearce2022ISPY}, with three important differences. First, we have an extra dependence on the edge profile, and hence the debris eccentricity $\eRMSI$; our predicted inner-edge location (Equation \ref{eq: rInAnalyticEstimate}) is increased by a factor of ${(1+2\eRMSI/\sqrt{3})}$ relative to that in \cite{Pearce2022ISPY}.

Second, \cite{Pearce2014} use a stricter definition of the sculpting timescale; they assume sculpting takes 10 diffusion times, which is the time it takes an eccentric planet to eject ${95\percent}$ of unstable material. However, we find that an edge sculpted by a circular-orbit planet with mass ${\mPOverMStar < 10^{-2}}$ takes just one diffusion timescale to assume roughly its final configuration (Equation \ref{eq: minMPMassFromDiffusionTime}). This is because it takes a planet exponentially longer to clear increasingly large fractions of unstable debris; a circular-orbit planet removes ${\sim70\percent}$ of unstable debris within 1 diffusion time, but takes 10 diffusion times to remove ${\sim90\percent}$ (Costa, Pearce \& Krivov, submitted). Hence our minimum-allowed planet mass can be a factor of ${\sqrt{10} \approx 3}$ times smaller than \cite{Pearce2022ISPY}.

Finally, we take the width of the chaotic zone around the planet's orbit to be 3 Hill radii, whilst \cite{Pearce2022ISPY} used 5. The reason for this is that \cite{Pearce2014} studied eccentric planets, for which the innermost stable semimajor axis is 5 eccentric Hill radii exterior to planet apocentre (their Figure 9), whilst for circular-orbit planets a value of 3 provides a better fit \citep{Gladman1993, Ida2000, Kirsh2009, Malhotra2021, Friebe2022}. It is unclear at what planet eccentricities the transition from 3 to 5 Hill radii occurs, though Figure \ref{fig: appEccentricPlanet} shows that a planet eccentricity of 0.2 is already high enough for 5 Hill radii to provide a better estimate. This is consistent with \cite{Regaly2018}, who find the transition occurs at planet eccentricities somewhere between 0 and 0.2 (their Figure 5c).

The result of these three differences is that any planets predicted using our model will differ slightly from the \cite{Pearce2022ISPY} model. However, the mass predictions should typically be within an order of magnitude or so of each other. For the example on Figure \ref{fig: examplePredictedPlanet}, the lower bound on planet mass predicted by the \cite{Pearce2022ISPY} model is 10 times smaller than the mass predicted from our model; the two predictions still lie in a similar region of parameters space, along Line 1 on that figure.

\subsection{Application to narrow discs}
\label{subsec: discussionNarrowDiscs}

We only considered broad discs in our simulations, to isolate the inner edge from the outer edge. This ensured that our inner-edge profiles were dominated by planet-disc interactions, rather than the initial profile of the disc. However, this means that our quantitative results may not hold for very narrow discs, where the outer- and inner-edge profiles could overlap.

In Section \ref{subsec: simulationsSDEccentricityOfInnerEdge} we argued that the characteristic inner-edge width is several times ${\rI \sigmaRI}$. Hence our results should hold if the disc is sufficiently wide for the edges to be well separated, i.e. 

\begin{equation}
    \rO - \rI \gtrsim \rI\sigmaRI + \rO \sigmaRO.
    \label{eq: narrowWidthCriterion}
\end{equation}

\noindent Equation \ref{eq: rSigInAnalyticEstimate} shows that $\sigmaRI$ is proportional to the eccentricity of inner-edge debris, and a similar relation holds for the outer edge. Therefore, if the eccentricity distribution were constant across the disc, then Equation \ref{eq: narrowWidthCriterion} implies that our results should hold provided the disc fractional width is larger than the rms debris eccentricity; this is in agreement with \cite{Marino2021}. Our full criterion (Equation \ref{eq: narrowWidthCriterion}) is slightly more complicated, because the debris-eccentricity level would vary across a planet-sculpted disc.

\subsection{Range of explored mass ratios}
\label{subsec: discussionRangeOfExploredPlanetMasses}

We explored planet-to-star-mass ratios ranging from ${3\times10^{-5}}$ to $10^{-1}$. We did not consider lower-mass planets, because in many cases the dynamical timescales would then become comparable to the stellar lifetimes. For example, an Earth-mass planet at ${10\au}$ from a Solar-type star (${\mPOverMStar=3\times 10^{-6}}$) would have a diffusion timescale of ${40\gyr}$, which would exceed the stellar lifetime of ${\sim10\gyr}$. Hence our results would not hold for very low mass ratios, because the interaction we study would not have time to occur.

Our upper limit of ${\mPOverMStar= 10^{-1}}$ represents the upper end of what could realistically be called a `planet'. For example, this could be a ${10\mJup}$ super-Jupiter orbiting a ${0.1\mSun}$ M6 dwarf. In many cases, such high-mass companions would be detectable via imaging, radial velocity or astrometry, particularly in younger systems (e.g. \citealt{Carter2023}).

Were a companion detected, then our results could be used to infer its evolutionary history. For example, if a debris disc in the system had an edge steepness consistent with sculpting by the observed companion, but the companion were detected far interior to the disc, then this could be evidence that the detected companion sculpted the disc historically but has since migrated inward. This scenario has been suggested as a way to explain non-detections of planets at the inner edges of debris discs \citep{Pearce2022ISPY}.

\section{Conclusions}
\label{sec: conclusions}

We perform a dynamical investigation into the effect of planets on the profiles of debris-disc inner edges. We consider both planet-disc interactions and debris collisions, and explore the interaction across a broad parameter space. We quantify our simulated surface-density profiles using an erf function, as in \cite{Rafikov2023}, for direct comparison with ALMA-resolved inner edges. Our main conclusions are as follows:

\begin{enumerate}
    \item For a non-migrating, circular-orbit planet, in the case where collisions are negligible, the steepness of the disc inner edge is set by the planet-to-star mass ratio and the initial-disc excitation level. Lower-mass planets lead to steeper inner edges, with the edge width proportional to ${(\mPOverMStar)^{1/3}}$ (Equations \ref{eq: planetERMSFit}, \ref{eq: eRMSFromPlanetAndInitialDisc} and \ref{eq: rSigInAnalyticEstimate}).

    \vspace{2mm}
    
    \item There is a maximum eccentricity that a planet on a circular, non-migrating orbit can generally impart on the population of planetesimals at a debris-disc inner edge, which is a root-mean-squared eccentricity of 0.06 (Equation \ref{eq: planetERMSFit}).

    \vspace{2mm}

    \item Considering the steepness of a debris disc's inner edge when inferring the properties of unseen sculpting planets can break the degeneracy between planet mass and semimajor axis. We provide a step-by-step method for inferring such planets from disc inner edges in Section \ref{subsec: modelExampleFitStepByStepMethod}. 

    \vspace{2mm}
    
    \item Eccentric planets make inner-edge profiles flatter, and introduce azimuthal asymmetries (Appendix \ref{app: eccentricPlanets}).

    \vspace{2mm}

    \item The inner edge of a purely planet-sculpted debris disc is much steeper than that of a purely collisional debris disc. 

    \vspace{2mm}

    \item Collisions flatten the profile of a planet-sculpted inner edge. The effect of collisions is small before the largest bodies start to collide; after this time, collisions start to significantly flatten the inner edge.

    \vspace{2mm}
    
    \item In most cases, collisions would never fully erase the signature of a sculpting planet on a debris-disc inner edge.

    \vspace{2mm}
    
    \item Whilst the inner edges of many ALMA-resolved debris discs are \textit{too steep} to be caused by collisions alone, they are also \textit{too flat} to arise through pure sculpting by non-migrating, circular-orbit planets, unless sculpting is still ongoing. Possible implications of this for the outer regions of planetary systems are discussed in Section \ref{subsec: discussionWhyObsEdgesFlatterThanSims}.
    
\end{enumerate}

\section*{Acknowledgements}

We thank the anonymous referee, whose comments improved the paper. We are grateful for exploratory work by Zuzanna Jonczyk as a summer intern at the Institute of Astronomy, University of Cambridge; her work helped identify some of the challenges in connecting inner-edge profiles to sculpting planets. We also thank Brenda Matthews for useful discussions. TDP, AVK, TL and MB were supported by Deutsche Forschungsgemeinschaft (DFG) grants \mbox{Kr 2164/13-2}, \mbox{Kr 2164/14-2}, \mbox{Kr 2164/15-2} and \mbox{Lo 1715/2-2}. TDP is also supported by a Warwick Prize Fellowship, made possible by a generous philanthropic donation. MB also received funding from the European Union’s Horizon 2020 research and innovation programme under grant agreement No. 951815 (AtLAST). AAS acknowledges support by the Alexander von Humboldt Foundation through a Humboldt Research Fellowship for postdoctoral researchers. MRJ acknowledges support from the European Union's Horizon Europe programme underthe Marie Sklodowska-Curie grant agreement No. 101064124, and funding provided by the Institute of Physics Belgrade, through the grant by the Ministry of Science, Technological Development, and Innovations of the Republic of Serbia. SM is supported by a Royal Society University Research Fellowship (URF-R1-221669).

\section*{Data Availability}

The data underlying this article will be shared upon reasonable request to the corresponding author.


\bibliographystyle{mnras}
\bibliography{bib} 

\appendix

\section{Surface-density profile fitting}
\label{app: fittingProcedure}

We fit surface-density profiles to our simulations through several rounds of $\chi^2$ minimisation, using the following process. First, we perform an initial $\chi^2$ minimisation, which typically provides a good fit to the broad disc as a whole. However, this generally produces poor fits at the edges, because the edges make up a small fraction of the overall profile; the inner edge on Figure \ref{fig: lowESurfaceDensities} spans just 25 radial bins, compared to the 400 bins across the entire disc, so even a poor edge fit would not significantly impact the overall $\chi^2$. To properly fit the inner edge, we then repeat the fitting using the parameters of the first fit as our initial guess, except for $\sigmaRI$ for which we use a value between $10^{-8}$ and $10^{2}$.  We repeat this several times for different initial guesses of $\sigmaRI$, and the fit with the lowest $\chi^2$ is taken as the new reference fit. Finally, this whole process is repeated for the outer edge, using the new reference fit as our initial guess and testing different guesses for $\sigmaRO$. The fit with the lowest $\chi^2$ after this process is the fit we use. This technique avoids local minima, and ensures that the edge profiles are properly fitted. It is also considerably faster than performing a full MCMC fit to each of our simulations. The fitting is performed using the {\sc python} module {\sc scipy.optimize}.

\section{Eccentric planets}
\label{app: eccentricPlanets}

We showed that a planet on a circular orbit excites debris eccentricities through MMRs, and that these eccentricities set the steepness of the disc inner edge. However, an eccentric planet would also drive up eccentricities through secular interactions. Whilst we focus on circular-orbit planets in this paper, in this section we briefly consider how planet eccentricity would affect the inner-edge profiles. We will show that a low- to moderate-eccentricity planet can excite debris more than a circular-orbit planet, resulting in a flatter inner edge, and that the planet eccentricities required to do this can be low enough that they do not necessarily impose a clear asymmetry on the disc.

\subsection{N-body simulation with an eccentric planet}
\label{subsec: appEccentricPlanetSimulation}

To demonstrate the effect of an eccentric planet, we re-run the simulation from Figure \ref{fig: lowESurfaceDensities}, but with the planet eccentricity increased to 0.2. We also initialise the disc inner edge to be slightly further outwards than before; we still place it one Hill radius exterior to the planet's apocentre, but since the planet is now eccentric, we switch to the definition of the eccentric Hill radius at apocentre from \cite{Pearce2014}:

\begin{equation}
    r_{\rm Hill, Q} \approx \aP(1+\eP) \left[ \frac{\mP}{(3-\eP) m_*} \right]^{1/3},
    \label{eq: eccentricHillRadiusAtApo}
\end{equation}

\noindent where $\eP$ is the planet eccentricity. We do this because the Hill radius of an eccentric planet varies around the planet orbit; the above equation is the Hill radius at the planet's apocentre, and the equivalent at planet pericentre is found by changing $\eP$ to $-\eP$ in this equation.

We run the eccentric-planet simulation for ${16.5\myr}$, which is much longer than the ${0.868 \myr}$ for the circular case. We do this because an eccentric planet would drive spiral density waves in a massless disc through secular interactions, in addition to sculpting the inner edge through scattering. \cite{Pearce2014} show that such a disc would settle into its final state after at least 10 secular times have elapsed at its outer edge (by which time the spirals are tightly wound and indistinct), \textit{and} at least 10 diffusion times have elapsed at the inner edge (by which time scattering would be largely complete). We compute the secular timescale using Equation 17 in \cite{Pearce2014}, and run our simulation to 10 times that value.

\begin{figure*}
	\includegraphics[width=17cm]{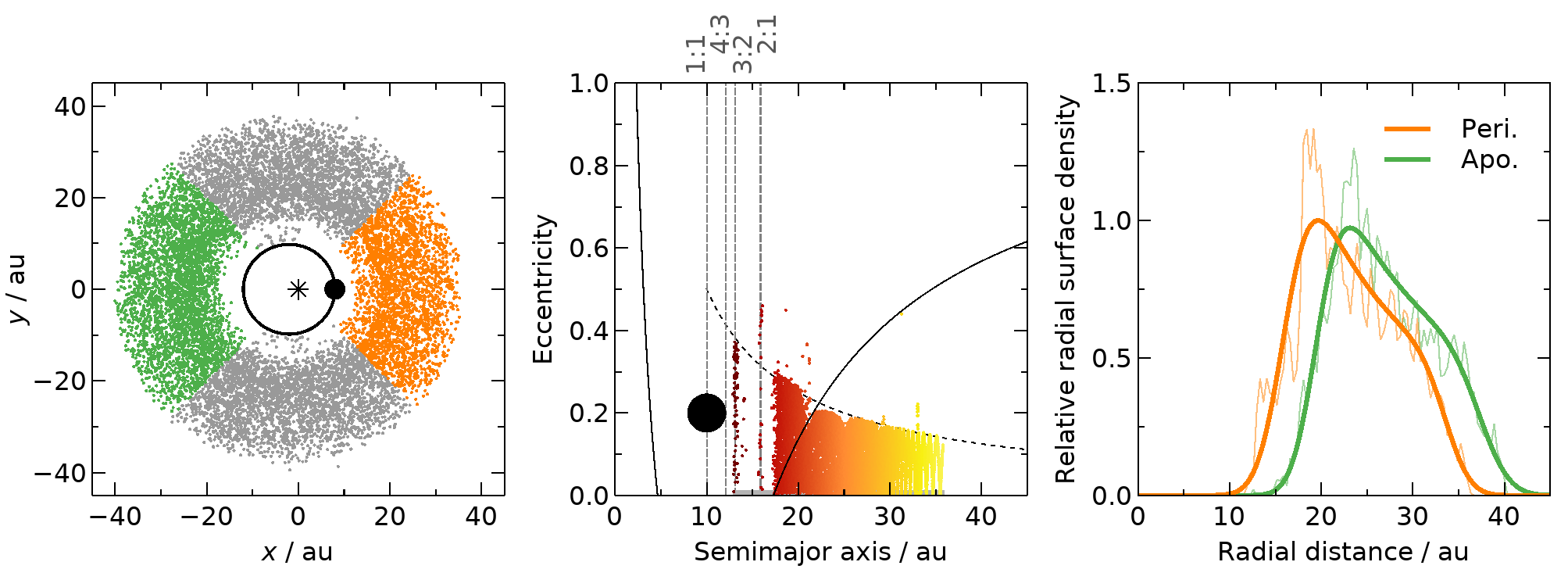}
    \caption{Simulation with an eccentric planet. The simulation has the same setup as that on Figure \ref{fig: lowESurfaceDensities}, except here the planet has eccentricity 0.2, and the initial disc lies further out (see Section \ref{subsec: appEccentricPlanetSimulation}). The eccentric planet excites debris through secular interactions, resulting in higher eccentricities and flatter edges than the circular-planet case. Left panel: positions at the end of the simulation. Orange and green points are debris within $45^\circ$ of the planet's pericentre and apocentre directions respectively. Middle panel: the dotted black line is ${2 e_{\rm forced}(a)}$, the maximum debris eccentricity expected from secular interactions (Equation \ref{eq: twoEForced}). The solid lines define the chaotic region, which is 5 (rather than 3) times the eccentric Hill radius either side of the eccentric planet's orbit. Right panel: separate surface-density profiles for material within $45^\circ$ of the planet's pericentre and apocentre. The thick and thin lines are the simulation data and fitted models respectively, relative to the peak of the pericentre model. The fit to the pericentre side has ${\rI=16.6\au}$, ${\sigmaRI=0.137}$, ${\rO=33.9\au}$ and ${\sigmaRO=0.0625}$, whilst that to the apocentre side has ${\rI=20.0\au}$, ${\sigmaRI=0.106}$, ${\rO=37.7\au}$ and ${\sigmaRO=0.0666}$; the peak fitted surface density on the apocentre side is 0.973 times that on the pericentre side. All remaining lines and symbols are defined on previous figures.}
    \label{fig: appEccentricPlanet}
\end{figure*}

Figure \ref{fig: appEccentricPlanet} shows the simulation with an eccentric planet, for comparison with the equivalent circular-planet simulation of Figure \ref{fig: lowESurfaceDensities}. As well as ejecting debris, the eccentric planet drives the disc into a broad, eccentric structure aligned with the planet orbit, as described in \cite{Pearce2014} and \cite{Faramaz2014}. Debris eccentricities are much higher than in the circular-planet case (Figure \ref{fig: appEccentricPlanet} middle panel, compared to Figure \ref{fig: lowESurfaceDensities} left panel), with implications for the inner-edge profile as discussed below. Another difference is that the inner edge is truncated at a semimajor axis corresponding to 5 eccentric Hill radii outside the planet's apocentre, rather than 3 Hill radii as in the circular case; this behaviour is discussed in \cite{Pearce2014}. The eccentric-planet simulation also has resonant debris surviving on planet-crossing orbits (e.g the 3:2 MMR), and such populations can be stable even for highly eccentric planets \citep{Pearce2021}.

We now assess the profile of the eccentric-disc inner edge. In our circular-planet simulations we simply azimuthally averaged the entire disc to produce radial surface-density profiles, but this approach is insufficient for asymmetric discs. Instead, we follow \cite{Tabeshian2016, Tabeshian2017} and divide the disc into several sectors. We define one sector as everything ${45^\circ}$ either side of the planet's pericentre direction, and another as everything ${45^\circ}$ either side of the planet's apocentre. These sectors are coloured orange and green respectively on the left panel of Figure \ref{fig: appEccentricPlanet}. For each of these sectors we azimuthally average the particles within them, to produce one radial surface-density profile for the region around pericentre and another for the region around apocentre. These profiles are shown on the right panel of Figure \ref{fig: appEccentricPlanet}. We fit each profile with Equation \ref{eq: fittedProfile} as before, again fixing the powerlaw index $\alphaR$ to 1.5. The result is that the inner edge has a different steepness on the apocentre and pericentre sides of the disc; we fit ${\sigmaRI=0.137}$ on the pericentre side, compared to 0.106 on the apocentre side. Both are much flatter than the circular-planet case, which had ${\sigmaRI=0.0361}$. In the next section we quantify the general impact of planet eccentricity on the inner-edge profile.

\subsection{Effect of planet eccentricity on the inner-edge profile}
\label{subsec: eccentricPlanetQuantify}

The inner-edge steepness is set by the eccentricity of surviving debris. A circular-orbit planet excites debris eccentricities through MMRs, whilst an eccentric planet also excites eccentricities through secular interactions. If the secular excitation were greater than the MMR excitation, then the eccentric planet would produce a flatter edge than a circular-orbit planet. In this section we calculate how eccentric a planet can be before the inner-edge profile deviates from the circular-planet case, and the effect of an eccentric planet on the inner edge.

Secular interactions with an eccentric planet cause debris eccentricities to oscillate, whilst their semimajor axes remain constant. Specifically, debris initialised on a circular orbit with semimajor axis $a$ would oscillate in eccentricity between zero and twice the forcing eccentricity $e_{\rm forced}$, where

\begin{equation}
    2e_{\rm forced}(a) \approx \frac{5}{2} \frac{\aP}{a} \eP
    \label{eq: twoEForced}
\end{equation}

\noindent \citep{Murray1999}. This is shown on the middle panel of Figure \ref{fig: appEccentricPlanet}, where the dotted line is $2e_{\rm forced}$. We can use this to derive the maximum eccentricity of secular debris at the disc inner edge; the innermost stable semimajor axis is about 5 eccentric Hill radii beyond planet apocentre, i.e. ${a_{\rm in} \approx \aP(1+\eP)+5r_{\rm Hill, Q}}$ \citep{Pearce2014}, so we can use Equation \ref{eq: eccentricHillRadiusAtApo} to get the innermost stable semimajor axis, then substitute this into Equation \ref{eq: twoEForced}. The result is that, to first order in planet eccentricity, the maximum debris eccentricity from secular effects at the disc inner edge is

\begin{equation}
    \eMaxSec \approx 2e_{\rm forced}(a_{\rm in}) \approx \frac{5}{2} \eP \left[1+3.47 \left(\frac{\mP}{m_*}\right)^{1/3} \right]^{-1}.
    \label{eq: eMaxSec}
\end{equation}

\noindent For the scenario on Figure \ref{fig: appEccentricPlanet} (${\eP=0.2}$, ${\mPOverMStar=1.9\times10^{-3}}$), Equation \ref{eq: eMaxSec} yields ${\eMaxSec \approx 0.35}$, in good agreement with the simulation.

We can use Equation \ref{eq: eMaxSec} to determine how eccentric a planet must be for the inner-edge profile to differ significantly from the circular-planet case. Equation \ref{eq: planetERMSFit} gives the rms eccentricity at the inner edge arising from MMRs from a circular-orbit planet; if an eccentric planet excited debris to higher values through secular interactions, then the edge profile would be flatter than in the circular-planet case. Comparing Equations \ref{eq: planetERMSFit} and \ref{eq: eMaxSec}, the disc inner edge would be flatter than the circular-planet case if the planet has eccentricity 

\begin{equation}
    \eP > 0.277 \left(\frac{\mP}{m_*}\right)^{1/3} + 0.961 \left(\frac{\mP}{m_*}\right)^{2/3}.
    \label{eq: minEpToAffectDisc}
\end{equation}

\noindent For the examples on Figures \ref{fig: lowESurfaceDensities} and \ref{fig: appEccentricPlanet} with a ${2\mJup}$ planet orbiting a solar-type star, ${\mPOverMStar=1.9\times10^{-3}}$ and so Equation \ref{eq: minEpToAffectDisc} predicts a critical planet eccentricity of 0.05 (equal to the eccentricity of Jupiter). So for this example, if the planet eccentricity were below 0.05 then the inner-edge profile would be similar to the circular-planet case (Figure \ref{fig: lowESurfaceDensities}), whilst higher planet eccentricities would result in a flatter inner edge (Figure \ref{fig: appEccentricPlanet}).

The actual inner-edge steepnesses on the pericentre and apocentre sides are comparable to that predicted by Equation \ref{eq: rSigInAnalyticEstimate}; for the example on Figure \ref{fig: appEccentricPlanet}, inserting ${\eRMSI\approx\eMaxSec/\sqrt{3} = 0.17}$ into \mbox{Equation \ref{eq: rSigInAnalyticEstimate}} predicts ${\sigmaRI\approx0.144}$, comparable to the simulation values of 0.137 and 0.106 on the pericentre and apocentre sides respectively\footnote{Strictly the rms eccentricity of a secular population scales with $\eMaxSec$ by a factor $\sqrt{2}$ rather than $\sqrt{3}$, since the eccentricity distribution is not uniform, but for this simple estimate we neglect this difference.}. 

Generally, the inner edge of an eccentric disc would be steeper on the apocentre side than the pericentre side, due to how the innermost debris evolves. \cite{Pearce2014} showed that the inner edge is set by debris with the innermost stable semimajor axis ${a_{\rm in}}$, which oscillates in eccentricity between 0 and $2e_{\rm forced}$. This debris also evolves in orientation, being maximally aligned with the planet orbit when its eccentricity is high, and minimally aligned when low. The result is that the disc inner edge assumes an eccentric shape; combining Equations 5 and 6 in \cite{Pearce2014} shows that an ellipse fitted to the disc inner edge will have eccentricity

\begin{equation}
    e_{\rm i} = \frac{e_{\rm forced}(a_{\rm in})}{1-e_{\rm forced}(a_{\rm in})},
    \label{eq: eInnerEdge}
\end{equation}

\noindent which is ${e_{\rm forced}(a_{\rm in})}$ to first order\footnote{The equivalent equation for the eccentricity of the outer edge is ${e_{\rm out} = e_{\rm forced}(a_{\rm out})/[1+e_{\rm forced}(a_{\rm out})]}$, where $a_{\rm out}$ is the outermost semimajor axis; similarly, this reduces to ${e_{\rm out} \approx e_{\rm forced}(a_{\rm out})}$ to first order.}. This means that at apocentre the inner edge is a superposition of low-eccentricity orbits at the innermost stable semimajor axis, so its profile is relatively sharp; conversely, at pericentre the inner edge is a diffuse superposition of higher-eccentricity orbits and is flatter. There may also be additional surface-density features that would manifest in particularly narrow and/or eccentric discs \citep{Pearce2014}.

\section{Criterion for planet-sculpted edges to be steeper than the overall disc}
\label{app: checkEdgeSteeperThanDisc}

In Appendix \ref{appSub: relatingErfToPowerlaw} we show that the steepness of the erf function is equivalent to that of a powerlaw $r^p$ if

\begin{equation}
    \sigmaRI = \sqrt{\frac{2}{\pi}} \frac{1}{p},
    \label{eq: sigmaRIToBeSteeperThanPowerlaw}
\end{equation}

\noindent and so the inner edge is much steeper than the overall disc profile if ${\sigmaRI \ll \sqrt{2/ \pi} / \alphaR}$. Since all of our simulated inner edges have ${\sigmaRI < 0.2}$ (Figure \ref{fig: sigmaRIn}), the planet-sculpted edges should be much steeper than the overall disc provided that the initial-disc profile is flatter than ${r^{-4}}$ (and also ${r^{4}}$).

\section{Deriving inner-edge parameters from our simple scattering model}
\label{app: simpleModelDerivations}

Here we use our simple scattering model to predict inner-edge parameters, yielding the results in Sections \ref{subsec: modelInnerEdgeLocation} and \ref{subsec: modelInnerEdgeSteepness}. These parameters are derived with reference to Figure \ref{fig: scatteringModel}.

\subsection{Semimajor axes $a_1$ and $a_2$, and the fraction of surviving debris with semimajor axis $a$}
\label{subsec: appA1AndA2}

We define

\begin{equation}
    a_1 \equiv \aP + \delta a
    \label{eq: a1}
\end{equation}

\noindent to be the semimajor axis where ${e_{\rm q}(a_1) = 0}$, i.e. the semimajor axis of a circular orbit at the outer edge of the chaotic zone (recalling that $e_{\rm q}(a)$ is given by Equation \ref{eq: eqFromPeriAtOuterEdgeOfChaoticZone}). Then we define

\begin{equation}
    a_2 \equiv \frac{\aP + \delta a}{1-e_{\rm max}} = \frac{a_1}{1-e_{\rm max}}
    \label{eq: a2}
\end{equation}

\noindent as the semimajor axis where ${e_{\rm q} = e_{\rm max}}$, i.e. the semimajor axis of an orbit with pericentre at the outer edge of the chaotic zone and eccentricity equal to the maximum eccentricity $e_{\rm max}$.

Next, we assume that the fraction of surviving particles with semimajor axis $a$ transitions from ${0\percent}$ for ${a \leq a_1}$ to ${100\percent}$ for ${a \geq a_2}$. Since we assume that all debris with initial eccentricity above ${e_{\rm q}(a)}$ is ejected, for particles with initial eccentricities uniformly distributed between 0 and $e_{\rm max}$ the probability density function of surviving particles goes as

\begin{equation}
    S(a) \propto \frac{e_{\rm q}(a)}{e_{\rm max}}
    \label{eq: sA}
\end{equation}

\noindent for ${a_1 \leq a \leq a_2}$. Similarly, ${S(a)=0}$ for ${a \leq a_1}$ and is constant for ${a \geq a_2}$.

\subsection{Inner-edge location $\rI$}
\label{subsec: appRIDerivation}

For an orbit with semimajor axis $a$ and eccentricity $e$, the time-averaged value of some parameter $x$ over a single orbit is 

\begin{equation}
    \langle x \rangle = \frac{1}{2 \pi a^2 \sqrt{1-e^2}} \int^{2\pi}_0 r^2(f) x(f) {\rm{d}}f ,
    \label{eq: timeAveragedParOverOrbit}
\end{equation}

\noindent where $r$ and $f$ are the radial distance and true anomaly respectively (e.g. \citealt{Murray1999}). Hence the time-averaged radial distance of a single body is

\begin{equation}
    \langle r_{a,e} \rangle = a\left(1 + \frac{e^2}{2} \right) .
    \label{eq: timeAveragedRadiusOverOrbit}
\end{equation}

\noindent For a collection of bodies with the same semimajor axis but with eccentricities uniformly drawn between 0 and $e_{\rm q}$, we can integrate Equation \ref{eq: timeAveragedRadiusOverOrbit} to get the average radial position of the bodies in this group; this is ${\langle r_{a} \rangle = \int^{e_{\rm q}}_{0} \langle r_{a,e}\rangle (e) {\rm d}e / e_{\rm q}}$, which yields

\begin{equation}
    \langle r_{a} \rangle = a\left[1 + \frac{e_{\rm q}^2 (a)}{6} \right] .
    \label{eq: averageRadiusAtA}
\end{equation}

Next, we integrate Equation \ref{eq: averageRadiusAtA} to find the average position of all particles with semimajor axes between $a_1$ and $a_2$, which we hypothesise to be similar to the characteristic inner-edge radius $r_{\rm i}$. This is ${\rI \approx \int^{a_2}_{a_1} \langle r_{a} \rangle (a) S(a) {\rm d}a}$. Finally, we replace $e_{\rm max}$ in the above equations with ${\sqrt{3} \eRMSI}$ for a uniform eccentricity distribution, and expand to first order in eccentricity. This yields our prediction for the characteristic radius of the sculpted disc's inner edge:

\begin{equation}
    \rI \approx \aP \left(1 + \frac{\delta a}{\aP} \right) \left(1+\frac{2}{\sqrt{3}} \eRMSI \right).
    \label{eq: rInAnalyticEstimateAppendix}
\end{equation}

\section{Relating the various literature models used to quantify edge profiles}
\label{app: relatingDifferenentEdgeProfileModels}

\begin{figure}
	\includegraphics[width=7cm]{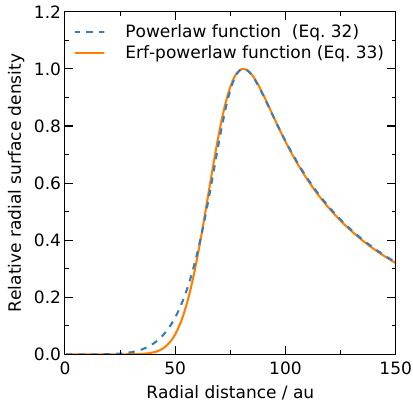}
    \caption{The double-powerlaw model for the observed \mbox{q$^1$ Eri} disc from \citet{Lovell2021}, compared to our erf-powerlaw model inferred using the method in Section \ref{sec: applicationToObservedDiscs}.}
    \label{fig: q1EriModelComparison}
\end{figure}

Various parametric models are used in the literature to quantify the steepness of debris-disc inner edges. Here we provide conversions of commonly used steepness parameters to the $\sigmaRI$ we use. We also show a radial profile quantified by a powerlaw function, and also the equivalent erf-powerlaw function with parameters derived using the following equations, to show the two are similar (Figure \ref{fig: q1EriModelComparison}).

\subsection{Relating erf to a radial powerlaw}
\label{appSub: relatingErfToPowerlaw}

Consider an inner edge fitted with the erf function

\begin{equation}
    \Sigma_{\rm E}(r) = \Sigma_0 \left[ 1-{\rm erf}\left(\frac{\rI - r}{\sqrt{2} \sigmaRI \rI} \right) \right],
    \label{eq: appErf}
\end{equation}

\noindent where ${\Sigma_0}$ is the surface density at the characteristic edge radius $\rI$ (e.g. \citealt{Rafikov2023}). The profile could alternatively be quantified by a powerlaw:

\begin{equation}
    \Sigma_{\rm P}(r) = \Sigma_0 \left(\frac{r}{\rI}\right)^{\pI}.
    \label{eq: appPowerlaw}
\end{equation}

\noindent Equations \ref{eq: appErf} and \ref{eq: appPowerlaw} can be differentiated to yield the slopes at $\rI$:

\begin{equation}
    \frac{\rm d}{{\rm d}r} \Sigma_{\rm E}(r)\bigg\rvert_{r=\rI} = \sqrt{\frac{2}{\pi}} \frac{\Sigma_0}{\rI \sigmaRI},
    \label{eq: appDErfByDr}
\end{equation}

\noindent and

\begin{equation}
    \frac{\rm d}{{\rm d}r} \Sigma_{\rm P}(r)\bigg\rvert_{r=\rI} = \frac{\Sigma_0 \pI}{\rI}.
    \label{eq: appDPowerlawByDr}
\end{equation}

\noindent Hence the erf function (Equation \ref{eq: appErf}) has the same steepness as the powerlaw function (Equation \ref{eq: appPowerlaw}) at $\rI$ if

\begin{equation}
    \sigmaRI = \sqrt{\frac{2}{\pi}} \frac{1}{\pI}.
    \label{eq: appSigmaRIEqualsPowerlawSteepness}
\end{equation}

\subsection{Relating erf to tanh}
\label{appSub: relatingErfToTanh}

Consider an inner edge fitted with a hyperbolic-tangent function

\begin{equation}
    \Sigma_{\rm T}(r) = \Sigma_0 \left[1+\tanh\left(\frac{r-\rI}{l_{\rm i}}\right) \right],
    \label{eq: appTanhEdgeFit}
\end{equation}

\noindent where $l_{\rm i}$ characterises the edge flatness (e.g. \citealt{Marino2021}). Differentiating this yields the slope at $r=\rI$:

\begin{equation}
    \frac{\rm d}{{\rm d}r} \Sigma_{\rm T}(r)\bigg\rvert_{r=\rI} = \frac{\Sigma_0}{l_{\rm i}}.
    \label{eq: appDTanhByDr}
\end{equation}

\noindent Equating this to Equation \ref{eq: appDErfByDr} shows that the erf function has the same slope as the hyperbolic-tangent function at $r=\rI$ if

\begin{equation}
    \sigmaRI = \sqrt{\frac{2}{\pi}} \frac{l_{\rm i}}{\rI}.
    \label{eq: appSigmaRIEqualsTanhSteepness}
\end{equation}

\subsection{Relating tanh to a radial powerlaw}
\label{appSub: relatingTanhToPowerlaw}

Equating Equations \ref{eq: appDPowerlawByDr} and \ref{eq: appDTanhByDr} shows that a radial powerlaw has the same slope as the hyperbolic-tangent function at $r=\rI$ if

\begin{equation}
    \pI = \frac{\rI}{l_{\rm i}}.
    \label{eq: appPowerlawSteepnessEqualsTanhSteepness}
\end{equation}

\bsp	
\label{lastpage}
\end{document}